\documentclass{article}

\usepackage{frontmatter/arxiv}
\usepackage[utf8]{inputenc} 
\usepackage[T1]{fontenc}    
\usepackage{hyperref}       
\usepackage{url}            
\usepackage{booktabs}       
\usepackage{amsfonts}       
\usepackage{mathrsfs}       
\usepackage{stmaryrd}
\usepackage{nicefrac}       
\usepackage{microtype}      
\usepackage{float}
\usepackage{graphicx}
\usepackage{natbib}
\usepackage{subcaption}
\usepackage{tikz}						

\usetikzlibrary{positioning,
               arrows,shapes,snakes,
		       automata,backgrounds,
		       petri,topaths}				

\tikzstyle{every draw}=[thick]

\tikzset{
    dotCircle/.style={circle, draw, thick, fill, inner sep=2pt},
    dotSquare/.style={draw, thick, fill, inner sep=2.5pt},
    dotDiamond/.style={draw, rotate=45, thick, fill, inner sep=2.5pt},
    rrec/.style={draw=black, rounded corners, text centered, node distance=1.5cm,
        minimum width=.6cm, minimum height=.6cm, thick},
    border/.style={dash pattern=on 3pt off 3pt, color=gray!70, thin},
    dist/.style={dash pattern=on 10pt off 2pt, thin},
    rel/.style={},
}

\newcounter{resultCounter}
\newcommand{\result}[1]{\refstepcounter{resultCounter}\paragraph*{Result \arabic{resultCounter}} \textit{#1}\\}

\newcommand{\discussion}[1]{\paragraph*{#1}}

\usepackage[acronym,toc,shortcuts]{glossaries}
\glsdisablehyper
\makeglossaries

\newacronym{arpe}{ARPE}{Année de Recherche Pré-doctorale à l’Étranger}
\newacronym{ens}{ENS}{École Normale Supérieure}

\newacronym{psp}{PSP}{postsynaptic potential}
\newacronym{ap}{AP}{action potential}

\newacronym{eeg}{EEG}{electroencephalography}
\newacronym{meg}{MEG}{magnetoencephalography}
\newacronym{ecog}{ECoG}{electrocorticography}
\newacronym{ieeg}{iEEG}{intracranial electroencephalography}
\newacronym{fmri}{fMRI}{Functional Magnetic Resonance Imaging}
\newacronym{pet}{PET}{Positron Emission Tomography}
\newacronym{spect}{SPECT}{Single Photon Emission Computed Tomography}
\newacronym{lm}{LM}{Light Microscopy}
\newacronym{em}{EM}{Electron Microscopy}
\newacronym{lfp}{LFP}{local field potential}

\newacronym{bci}{BCI}{brain-computer interface}
\newacronym{mi}{MI}{mental imagery}
\newacronym{mi_bci}{MI-BCI}{mental imagery based brain-computer interface}
\newacronym{erp}{ERP}{event-related potential}
\newacronym{ssvep}{SSVEP}{steady-state visually evoked potential}
\newacronym{erf}{ERF}{event-related field}

\newacronym{ml}{ML}{machine learning}
\newacronym{dl}{DL}{deep learning}
\newacronym{knn}{KNN}{K-Nearest Neighbors}
\newacronym{pca}{PCA}{Principal Component Analysis}
\newacronym{tsne}{t-SNE}{t-distributed Stochastic Neighbor Embedding}
\newacronym{umap}{UMAP}{Uniform Manifold Approximation and Projection}
\newacronym{lda}{LDA}{Linear Discriminant Analysis}
\newacronym{nn}{NN}{neural network}
\newacronym{cnn}{CNN}{convolutional neural network}
\newacronym{lstm}{LSTM}{long short-term memory}
\newacronym{rnn}{RNN}{recurrent neural network}
\newacronym{elu}{ELU}{exponential linear unit}
\newacronym{grl}{GRL}{gradient reversal layer}
\newacronym{fbcsp}{FBCSP}{Filter Bank Common Spatial Pattern}

\newacronym{loso}{LOSO}{leave one subject out}

\newacronym{hgd}{HGD}{High-Gamma Dataset}

\newacronym{gpu}{GPU}{graphics processing unit}
\newacronym{cpu}{CPU}{central processing unit}

\newacronym{inria}{INRIA}{Institut National de Recherche en Informatique et en Automatique}
\newacronym{opal}{OPAL}{Observatoire Pluridisciplinaire des ALpes-maritimes}
\newacronym{uca}{UCA}{Université Côte d'Azur}

\newcommand{\anchor}{\mathrm{a}}
\newcommand{\positive}{\mathrm{p}}
\newcommand{\negative}{\mathrm{n}}
\newcommand{\rpz}{\mathnormal{f}}
\newcommand{\example}[1]{x_{#1}}
\newcommand{\embedvector}[1]{v_{#1}}

\newcommand{\exampleset}{\mathcal{X}}
\newcommand{\vectorset}{\mathcal{V}_\rpz}
\newcommand{\indices}[2]{\mathcal{I}_{#1}^{#2}}
\newcommand{\tripletsindices}{\mathcal{T}}
\newcommand{\distsymbol}{\mathnormal{d}}

\newcommand{\examplea}{\example{\anchor}}
\newcommand{\examplep}{\example{\positive}}
\newcommand{\examplen}{\example{\negative}}
\newcommand{\vectora}{\embedvector{\anchor}}
\newcommand{\vectorp}{\embedvector{\positive}}
\newcommand{\vectorn}{\embedvector{\negative}}
\newcommand{\indicesa}[1]{\indices{\anchor}{#1}}

\newcommand{\rawdist}[2]{\distsymbol\big(#1,#2\big)}

\newcommand{\vdist}[2]{\rawdist{\embedvector{#1}}{\embedvector{#2}}}
\newcommand{\vdista}[1]{\vdist{\anchor}{#1}}
\newcommand{\vdistap}{\vdista{\positive}}
\newcommand{\vdistan}{\vdista{\negative}}

\newcommand{\maxzero}[1]{\left[ #1 \right]_+}
\newcommand{\subjectt}{ \texttt{subject} }
\newcommand{\targett}{ \texttt{im\_class} }
\newcommand{\eqdef}{\buildrel \text{\tiny def}\over =}
\newcommand{\loss}{\mathscr{L}}

\newcommand{\intset}[1]{\llbracket #1 \rrbracket}

\makeatletter
\newcommand{\subalign}[1]{%
  \vcenter{%
    \Let@ \restore@math@cr \default@tag
    \baselineskip\fontdimen10 \scriptfont\tw@
    \advance\baselineskip\fontdimen12 \scriptfont\tw@
    \lineskip\thr@@\fontdimen8 \scriptfont\thr@@
    \lineskiplimit\lineskip
    \ialign{\hfil$\m@th\scriptstyle##$&$\m@th\scriptstyle{}##$\hfil\crcr
      #1\crcr
    }%
  }%
}
\makeatother

\newcommand{\embedFigure}[4]{
\begin{figure}[h!]
  \centering
  \includegraphics[width=\linewidth]{export/#1_subj#2}
  \caption[#3 visualization of the embedding space with #2 as test subject]{Visualization of the embedding spaces using #3 with subject #2 as test subject.
  Each dot is the projection of an embedding vector in 2D by #3.
  The brown dots with a black edge denote data points of the test subject #2. #4}
\end{figure}
}

\newcommand{\desca}{original triplet loss}
\newcommand{\descb}{weighted lexicographic order}
\newcommand{\descc}{lexicographic order}
\newcommand{\descd}{product order}

\usepackage{epstopdf}
\AtBeginDocument{%
  \PrependGraphicsExtensions*{
    .eps,.EPS,.ps,.PS,.pdf,.PDF,
    .png,.PNG,.jpg,.jpeg,.JPG,.JPEG
  }%
}

\title{An embedding for EEG signals learned using a triplet loss}

\date{October 3, 2021}	

\author{
	\href{https://orcid.org/0000-0002-8933-7640}{\includegraphics[scale=0.06]{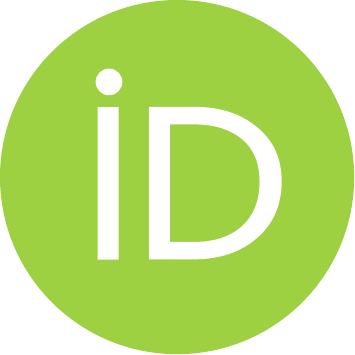}\hspace{1mm}Pierre Guetschel} \\
	ENS Paris-Saclay \\
  France \\
	\texttt{pierre.guetschel@ens-paris-saclay.fr} \\
	\And
	\href{https://orcid.org/0000-0002-1643-9988}{\includegraphics[scale=0.06]{logos/orcid.pdf}\hspace{1mm}Théodore Papadopoulo} \\
	Université Côte d’Azur, INRIA \\
	France \\
	\texttt{Theodore.Papadopoulo@inria.fr} \\
	\AND
	\href{https://orcid.org/0000-0001-6729-0290}{\includegraphics[scale=0.06]{logos/orcid.pdf}\hspace{1mm}Michael Tangermann} \\
	Donders Institute for Brain, Cognition and Behaviour \\
	Radboud University \\
	Nijmegen, The Netherlands \\
	\texttt{michael.tangermann@donders.ru.nl} \\
}

\hypersetup{
pdftitle={EEG Embeddings},
pdfsubject={cs.HC},
pdfauthor={Pierre Guetschel, Théodore Papadopoulo, Michael Tangermann},
pdfkeywords={EEG, embedding, deep learning, metric learning, triplet loss, ladder loss, calibration},
}

\usepackage[normalem]{ulem}

\begin{document}
\maketitle

\begin{abstract}

Neurophysiological time series recordings like the electroencephalogram (EEG) or local
field potentials are obtained from multiple sensors.
They can be decoded by machine learning models in order to estimate
the ongoing brain state of a patient or healthy user.
In a brain-computer interface (BCI), this decoded brain state information can be used with minimal time delay to either control an application, e.g., for communication or for rehabilitation after stroke,
or to passively monitor the ongoing brain state of the subject, e.g., in a demanding work environment.
A specific challenge in such decoding tasks is posed by the small dataset sizes in BCI compared to other domains of machine learning like computer vision or natural language processing.
A possibility to tackle classification or regression problems in BCI despite small training data sets is through transfer learning, which utilizes data from other sessions, subjects or even datasets to train a model.
\\
In this exploratory study, we propose novel
domain-specific embeddings for neurophysiological data.
Our approach is based on metric learning and builds upon the recently proposed ladder loss.
Using embeddings allowed us to benefit, both from the good generalisation abilities and robustness of deep learning and from the fast training of classical machine learning models for subject-specific calibration.
In offline analyses using EEG data of 14 subjects, we tested the embeddings' feasibility and compared their efficiency with state-of-the-art deep learning models and conventional machine learning pipelines. For within-subject classification of motor tasks, our data suggests that  embeddings significantly improve the decoding accuracy compared to an established baseline model. In between-subjects transfer scenarios, pre-trained embeddings allow for an extremely fast fine-tuning of a classifier for a novel subject, even if only a few trials per class are available, which currently can not be achieved with regular deep learning models.
And in the scenarios where one subject is completely set aside for testing without specific calibration, the embeddings were not competitive.
\\
In summary, we propose the use of metric learning to obtain pre-trained embeddings of EEG-BCI data as a means to incorporate domain knowledge and to reach competitive performance on novel subjects with minimal calibration requirements.

\end{abstract}
\section{Introduction}

The activity of the brain can be monitored by different invasive or non-invasive methods
like electroencephalography (EEG) or local field potential (LFP).
The classification of the associated signals by machine learning models
allows to estimate the ongoing state of the brain
and thus to build closed-loop neurotechnological systems like brain-computer interfaces (BCIs).
Through these interfaces, paralyzed patients can send commands to software,
enabling them to communicate ~\cite{guy2018brain}
or to control prosthesis~\cite{muller2007control}.
While brain-computer interface and corresponding experimental paradigms~\cite{abiri2019comprehensive} originally have been designed for individuals with disabilities~\cite{moghimi2013review}, non-medical applications~\cite{van2012brain}
and specifically BCI-based gaming~\cite{tangermann2008playing,ahn2014review,vasiljevic2020brain} has been explored. 
In neuroscience and in BCI, datasets are very small and usually contain at most a few dozen different subjects.
This is all the more important as brain signal are notably variable across subjects and sessions.
Traditional BCI approaches with machine learning pipelines based on linear classification and the learning of individual informative subspaces can cope with this challenge, but small datasets unfortunately limit the possibility to exploit the recent advances in deep learning for the field of BCI.
Nonetheless, deep models could be used successfully to decode data obtained under different BCI paradigms, if the datasets were sufficiently large and if state-of-the-art regularization approaches were utilized.
Cecotti and Graser proposed a \acrfull{cnn} model to classify \acrfull{erp} data
and were the first to explore CNNs in BCI ~\cite{cecotti2010convolutional}.
The Riemannian geometry is commonly used in traditional brain-computer interfaces
~\cite{barachant2011multiclass,yger2016riemannian}
and is still used to this day~\cite{chu2020decoding}.
%
For an \acrfull{ssvep} task, Kwak et al.~
proposed a CNN architecture that outperformed most traditional approaches
on noisy signals~\cite{kwak2017convolutional}. 
Schirrmeister and colleagues classified EEG data obtained under motor execution tasks
with ShallowConvNet, DeepConvNet and other architectures
and emphasized the interpretability of their deep models~\cite{schirrmeister2017deep}.
The authors of EEGNet~\cite{lawhern2018eegnet} showed that their architecture was capable of classifying EEG recordings from four different BCI tasks. Compared to other architectures, EEGNet performs extremely competitive despite a low number of parameters and thus relatively short training times.
And recently, the FBCNet architecture was proposed by Mane and colleagues~\cite{mane2021fbcnet}. It was designed to imitate
the design of the successful classical EEG classification algorithm
called \acrshort{fbcsp}~\cite{ang2008filter} for data obtained under mental imagery tasks.
A panorama of the different machine learning algorithms
used in EEG-based brain-computer interfaces is given by ~\cite{lotte2019inria}
and a review focusing specifically on deep learning approaches has been provided by Roy and colleagues~\cite{roy2019deep}. 
For most approaches described in the literature, the standard within-subject decoding approach is used, i.e., to train one model per subject. This approach can be justified as brain signals are rather individual
and subjects may share only certain brain signal features when they perform the same task.
Cross-subject, cross-session and cross-dataset models are considered as extremely challenging, and common representations capable to relate data of multiple subjects or multiple sessions are still lacking.
Yet, training deep models is time-consuming and non-trivial to automatize: they
typically require a hyper-parameters search per subject and a careful monitoring of the training process often is necessary to reach convergence on a competitive decoding level.
Thus the existing subject-specific artificial neural networks can not realistically be trained under the time pressure of online studies, i.e., when the BCI user is waiting under the EEG cap to start using the online application.
For this reason, research on models capable of generalization is required.

Embeddings are rather common in the image and language domains, among others.
They are lower-dimensional spaces into which high-dimensional vectors can be projected.
Usually these embedding spaces come with a metric,
which expresses a notion of similarity between the embedded elements.
For example, in face embeddings two images obtained from the same person
are expected to be close together while the images of two different persons should be represented more distantly from each other in the same space.
While embeddings typically provide the benefit to reduce the dimensionality, e.g., those obtained by autoencoders, they are utilized also to represent data elements in a structured manner and to act as a pre-processing. If an embedding is versatile, it can be used as the basis for multiple machine learning tasks.
In computer vision, the face image embeddings are vectors
that represent picture of the faces of people ~\cite{schroff2015facenet}.
They allow for face verification (is this the same person?), recognition (who is this person?) and clustering.
With a well-generalized face embedding, a new person's face
can be differentiated from other faces without the need to retrain the
model that translates images into embedding vectors.
This is a significant advantage for the deployment of machine learning models in industrial applications.
Word embeddings are at the foundation of modern natural language processing (NLP)~\cite{alvarez2017review}.
They consist of mappings from vocabularies to vector spaces~\cite{goldberg2017neural}.
These vectors are to the machines what definitions in dictionaries are to humans -- they give meaning to words.
Making use of pre-trained word embeddings, NLP models are build on top of them. 
This allows NLP models to be trained on much smaller datasets than if raw word inputs were used. 
In addition, by using word embeddings, NLP models can generalize to words outside of their training corpus ~\cite{goldberg2017neural}. 
The authors of the Word2vec embedding~\cite{mikolov2013efficient} even observed
some unexpected structures, close to an algebra, that emerged in their embedding space:
for example $vector("King") - vector("Man") + vector("Woman")$ results in a vector
that is closest to the embedding of the word $Queen$. Even if these relations were never explicitly taught to the model, they were deduced from the use of these words in the training corpus.

Embeddings and representation learning refer to very similar concepts
and are sometimes used interchangingly in the literature.
In the following, we will assume that an \emph{embedding} denotes a function which maps input data points into a vector space.
This space is equipped with a metric where the distance between mapped vectors is supposed to approximate
a notion of similarity which had been observed for the original data elements (metric learning).
A \emph{representation} is a more permissive concept that only designates a transformation of the data. It typically is the side product of, e.g., a classification task, when different feature representations are observed from layers of a network after its training. So an embedding is also a representation but the converse is not true.
In the past year, a number of publications have explored representation learning
in the context of BCI and passive mental state monitoring.
Ko et al.~ developed a multi-scale \acrshort{cnn} to cope
for the multiple frequencies present in brain signals~\cite{ko2021multi}.
They validated their model by training it on data of different active and passive BCI paradigms, among them motor imagery tasks, \acrshort{ssvep} protocols, seizure data and vigilance estimates during driving.
On a motor imagery classification task, Jeon and colleagues proposed a representation based on mutual information maximization and minimization ~\cite{jeon2021mutual}.
For emotion recognition, Rayatdoost and colleagues learned a representation
invariant to subject or domain ~\cite{rayatdoost2021subject}
by using a \acrlong{grl}~\cite{ganin2016domain}.
Zhang and Etemad used an \acrshort{lstm} network (\acrlong{lstm}) 
trained on a regression task to predict attention levels during a driving task~\cite{zhang2021capsule}.
These representation learning papers have in common that
they obtained their representation as the output of one of the last layers of a deep architecture, which had been trained either on a classification or a regression task.
An exception is the approach by
Zhang and Etemad, who also trained an \acrshort{rnn}-based (\acrlong{rnn}) 
autoencoder on EEG signals to obtain a latent representation.
This representation was further on utilized to recognize emotions ~\cite{zhang2021deep}.

To our knowledge, so far only two papers proposed to make use of metric learning to obtain embeddings for brain signals.
The first is a recent paper by Alwasiti et al.~\cite{alwasiti2020motor}. The authors reported using the SoftPN loss~\cite{balntas2016pn}
for training deep architectures to generate embedding vectors
for a motor imagery dataset~\cite{goldberger2000physiobank}. The dataset consisted of the three classes left hand, right hand and rest and is rather small, containing around 120 trials per subject.
Feeding spectrograms to their networks, the SoftPN loss utilized triplets of embeddings, each consisting of two data points from the same class (positives) and one datapoint from another class (negative). The loss minimises the distance between the positives and maximises the distance between the negative and the closest positive, which is a relatively similar approach to the triplet loss we applied in our study.
For evaluation, Alwasity and colleagues classified the embedding vectors by a nearest neighbors model and reported the accuracy scores.
The authors reported the average scores of subject-specific embeddings,
a comparison between two spectrogram computing methods and the observation that cross-subject models performed worse than within-subject ones.
In the second study, Mishra and Bhavsar~\cite{mishra2021eeg} developed their own \acrshort{cnn} architecture and
used a dataset containing EEG recordings of participants that were shown pictures of either
characters (A-J), digits (0-9) or objects from 10 classes obtained from ImageNet~\cite{tirupattur2018thoughtviz}.
The study focused on three goals: separation of the three main classes (characters, digits and objects),
separation of the ten sub-classes (in three separate experiments)
and impact of EEG channel selection.
Metric learning was only involved to investigate the second goal: the authors
compared the performance obtained by their architecture
trained using a classification loss (probably the cross-entropy loss)
to that of their architecture trained using a triplet loss to embed the signal, and a k-nearest neighbour on top to classify the embedding vectors.
They showed that the triplet loss led to an improved score
over the classification loss and over losses reported by previous studies for the same datasets.

Inspired by their ability to generalize but also to cluster and to reveal unexpected features in the data,
we believe that embeddings based on metric learning have a strong potential to extract interesting structures from EEG recordings.
In our exploratory study, we examine different strategies to learn embeddings for EEG signals.
While the two existing studies~\cite{alwasiti2020motor,mishra2021eeg} have made a first step to propose such strategies, they did not leave any subject out during testing. Thus it is hard to assess the generalization ability of their models specifically to data of new subjects -- a transfer learning problem which is investigated intensively the field of BCI for obvious reasons.
We however believe that embeddings have the potential to significantly reduce calibration times in BCI applications.
The BCI transfer learning techniques with classical machine learning models . They are usually based on ensemble models ~\cite{zou2019inter} or sample weighting based on domain similarity ~\cite{zheng2021spatio}. But these techniques are usually used on domains with relatively close distributions, ie. domains coming from the same dataset.
Some deep learning techniques also exist for domain adaptation ~\cite{ko2021survey}, but they all imply (re)training a deep architecture on data from the target domain. This is not feasible for an online calibration while the subject is waiting.
Whereas with an embedding, one can train a deep architecture, on multiple data sources, to extract a powerful representation and train only a simple machine learning model, on few examples from the target domain, to classify these representations.
This paradigm will be seen in details in Section~\ref{sec:embed_eval}. It allows to exploit both the generalisation ability and the robustness of deep learning and the short target-specific calibration times of classical machine learning models.

We will propose a new variation of the triplet loss.
It allows to incorporate expert knowledge of the input domain to learn embeddings
and has strongly been inspired by the ladder loss
proposed for combined embeddings of images and text by Zhou and colleagues~\cite{zhou2020ladder}.
In offline analyses, we will explore different use cases and scenarios for these embeddings in the context of BCI,
including a simple subject-specific training,
a cross-subject training with minimal subject-specific calibration
and a pure cross-subject generalisation scenario.
Then, the use of these embeddings will be compared to state-of-the-art classification approaches in the different scenarios.
Finally, we will investigate if structures emerge from these embeddings.

\section{Materials and Methods}

This section first revisits the two state-of-the-art models EEGNet and \acrfull{fbcsp} against which our own models will be compared, and a widely used EEG dataset which allows for within-subject and across-subject training scenarios.

The following subsections will introduce triplet loss and two extensions which
will be used to learn embeddings.
Finally, we present details of the analyses and evaluation procedures,
including how embeddings are compared to other models and an overview of all considered models.


\subsection{Machine learning models and architectures}
\begin{figure}[h!]
    \centering
    \includegraphics[width=.5\textwidth]{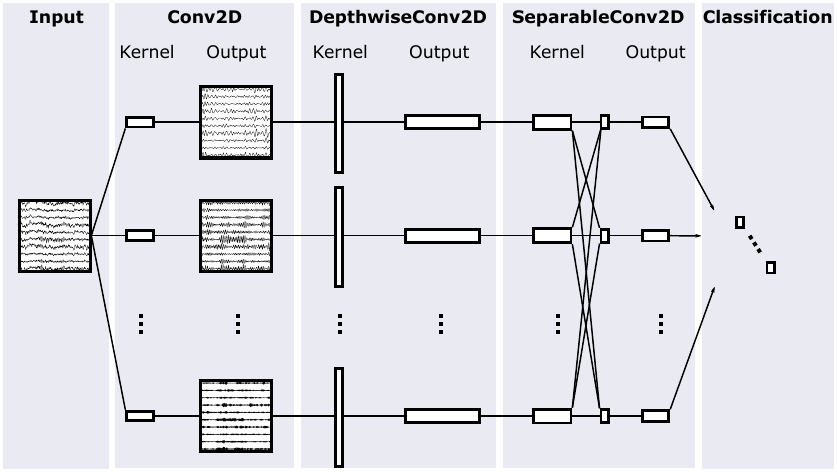}
    \caption[EEGNet architecture]{EEGNet architecture.
       Reproduced according to~\cite{lawhern2018eegnet}.}
       \label{fig:eegnet_arch}
\end{figure}

\paragraph*{EEGNet}
is a popular architecture that we will use in our experiments for multiple purposes.
It was originally designed by \cite{lawhern2018eegnet} as an end-to-end \acrshort{eeg} classifier.
Its authors successfully trained it on four different \acrshort{bci} classification tasks that include both \acrshort{erp} and oscillatory data.
The architecture of EEGNet feeds raw EEG input through first a temporal convolution to learn frequency filters
and second a depthwise convolutional layer to learn spatial features,
followed by a non-linearity and an average pooling layer,
before a separable convolutional layer
with non-linear activation and average pooling,
providing a high-level feature representation
for a final classification layer, as depicted in Figure~\ref{fig:eegnet_arch}.

In our study, we use EEGNet both in a classification configuration to serve as a baseline,
and in an embedder configuration.
The differences between these configurations are (1) the presence of a softmax output layer in the classifier but not in the embedder and (2) and the number of outputs.
In the classification configuration, EEGNet takes accepts raw EEG signals as inputs
and has as many outputs as there are classes, with every output estimating the probability that the input signal belongs to the corresponding class.
In the embedding configuration, EEGNet acts as the embedding function $\rpz(\cdot)$ (cf. Section~\ref{sec:tri_loss}).
Again, it takes raw EEG signals as input but outputs a non-normalized embedding vector with an arbitrary number of dimensions.
Based on our good prior experience in EEG-based BCI and in deep learning,
we decided to train EEGNet using the AdamW optimizer~\cite{loshchilov2017decoupled,kingma2014adam}
and with a 1cycle learning rate policy~\cite{smith2019super}.
The implementation we used is \texttt{EEGNetv4} from Braindecode~\cite{schirrmeister2017deep}
and we trained our models using PyTorch~\cite{paszke2019pytorch}.

\paragraph*{\acrfull{fbcsp}}\label{sec:fbcsp}
is a traditional EEG classification approach for imagery data~\cite{ang2008filter, chin2009multi}. It does not utilize a neural network. We use it as the second baseline method to compare our results against. The computation steps of \acrshort{fbcsp} are as follows: First, a \textbf{Bandpass filtering} is applied to obtain 9 non-overlapping frequency bands between 4\,Hz and 40\,Hz. Second, 4 spatial filters per frequency band are learned in a supervised approach using the Common Spatial Patterns (\textbf{CSP}) algorithm, which is widely applied for the extraction of oscillatory features~\cite{tangermann2012review}.
Third, applying these \textbf{spatial filters} to the bandpass filtered data delivers $9\times 4=36$ features per trial, which in a fourth step are classified by a \textbf{logistic regression classifier}.

\subsection{Dataset}\label{sec:dataset}
In this study, we train and compare our models on the publicly available \acrlong{hgd}~\cite{schirrmeister2017deep}.
It contains 128-channel \acrshort{eeg} recordings of fourteen healthy subjects executing three different movement classes and a resting condition. The movements are executed by the left hand, the right hand, and both feet.
During each trial, the subject is asked to execute one of the movement or a rest for a duration of four seconds.
The exact number of trials varied between subjects but was balanced between the four classes. For each subject, approximately 1000 trials were recorded in a single session structured into 13 runs.
The authors defined a training set formed by all but the last two runs, which contained approximately 880 trials per subject. We adopted this definition for our experiments. The remaining test set thus contains the 160 trails of the last two runs of each subject.

\begin{figure}[h!]
  \centering
  \includegraphics[width=.2\textwidth]{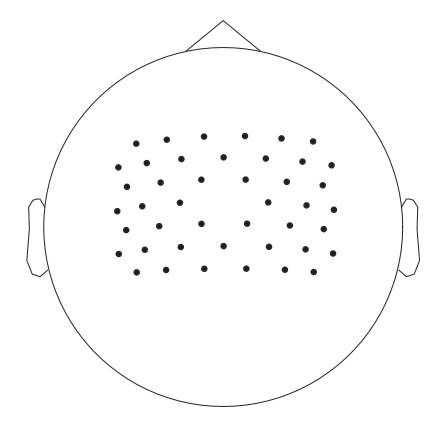}
  \caption[EEG sensors positions]{Position of the 44 EEG sensors used. Please note that the central channel Cz is removed, as it had been used as the reference channel of the recording.}
  \label{fig:sensors}
\end{figure}

In all our experiments, we only use a subset of 44 EEG sensors covering the motor cortex (as did the authors). Their location is shown in Figure~\ref{fig:sensors}.

Even though the higher gamma frequencies have been reported to contain utile information, the majority of studies conducted with this dataset focussed on the alpha and beta bands. In order to limit the network and dataset sizes for fast prototyping, we thus decided to low-pass filter the signal at 40\,Hz and downsample it to 128\,Hz, which is the same sampling frequency as the one used by the authors of EEGNet~\cite{lawhern2018eegnet}. This decision allowed us to use the EEGNet architecture unaltered for our experiments.
For all experiments, we corrected for drifts in the signal by applying, to each trial, a baseline correction that removes its average.
From each trial, we use the complete four seconds as input for the models. This corresponds to the complete time interval of movement execution, including possible delays due to reaction time.
This preprocessing leads to a representation of each trial in a matrix with dimensionality $512\times 44$.
All models are trained using this exact same pre-processing.
The dataset was downloaded and wrapped through the MOABB library~\cite{jayaram2018moabb}.

\subsection{Triplet loss and extensions}\label{sec:tri_loss}
In this work, we define and use a variation of the triplet loss
to train embeddings for EEG signals using neural networks.
We will compare different configurations of this new loss.

In the following, we represent the embedding function by $\rpz(\cdot)$.
It embeds an \acrshort{eeg} signal $\example{}\in\mathbb{R}^{\text{time}\times\text{channels}}$ into a $\mathrm{d}$-dimensional vector space $\mathbb{R}^\mathrm{d}$ which is equipped with a distance function $\rawdist{\cdot}{\cdot}$.
The embedding is learned such that this distance in the embedding space
expresses a notion of similarity between the \acrshort{eeg} signals.
In our work, $\rpz(\cdot)$ is computed by EEGNet and the distance function is considered fixed
and euclidean: $\forall \embedvector{1},\embedvector{2} \in \mathbb{R}^\mathrm{d}, \rawdist{\embedvector{1}}{\embedvector{2}} = \lVert{\embedvector{1}-\embedvector{2}}\rVert_2$.
We learn embedding spaces with eight dimensions, a parameter which had been determined by a preliminary experiment.
This choice is discussed in Section~\ref{sec:discussion}.
We define $\exampleset\eqdef\{\example{i}\}_{i\in\intset{1,N}}$ as the set of EEG trials
and $\vectorset\eqdef\{\embedvector{i}\}_{i\in\intset{1,N}}$ as the corresponding embedding vectors
such that $\forall i\in\intset{1,N}, \embedvector{i}\eqdef\rpz(\example{i})$.
To utilize EEGNet for the training of embeddings, we use different versions of a loss function called the triplet loss.

\begin{figure}[h!]
\centering
\includegraphics[width=\linewidth]{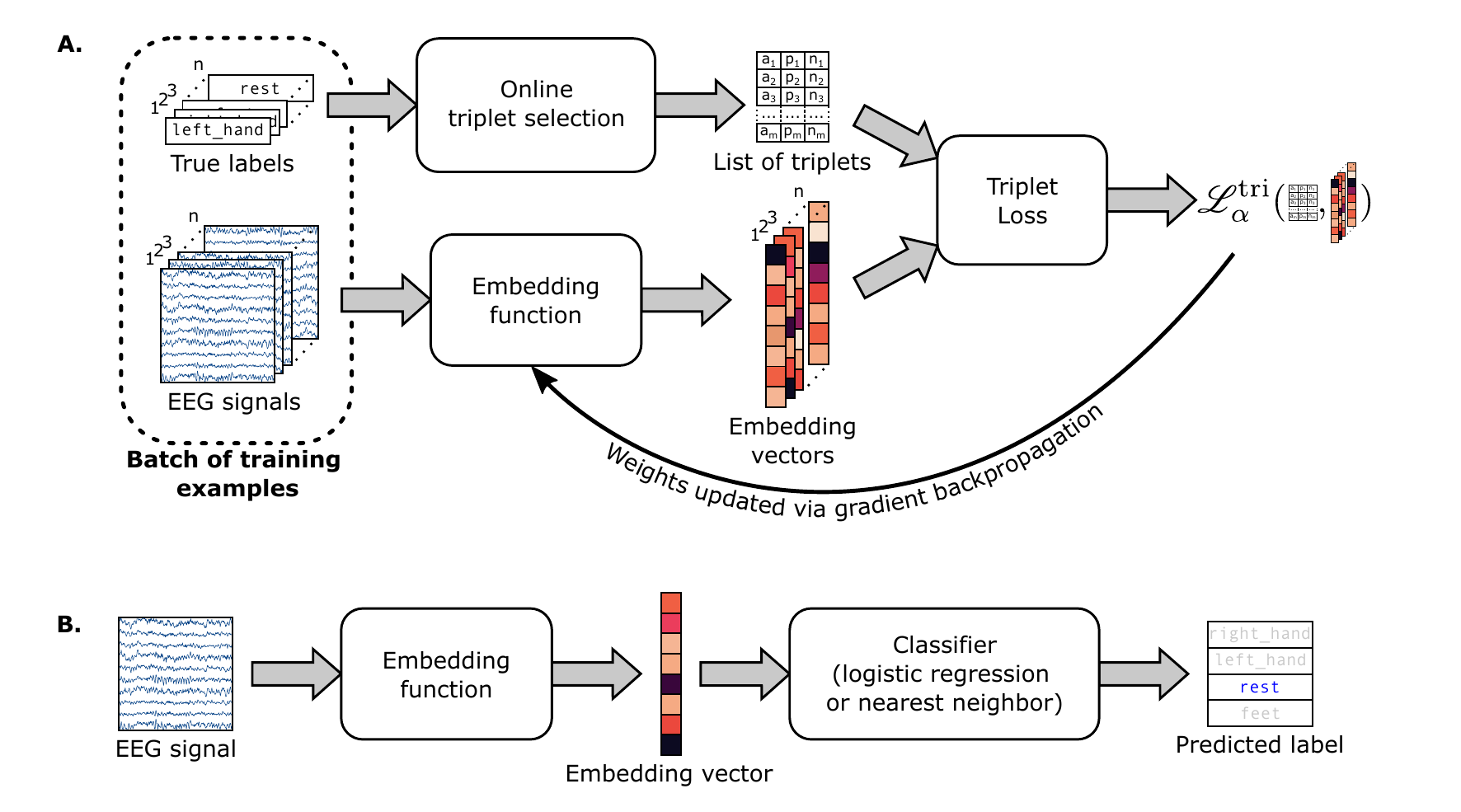}
\caption[Diagrams of embedding functions training and testing]{
  Diagram \textbf{A} describes the training procedure of an embedding function when using a triplet loss (cf.~Section~\ref{sec:tri}) and an online triplet selection (cf.~Section~\ref{sec:triplet_select}). At every training iteration, the algorithm takes as input a batch of EEG trials along with their respective labels. The EEG recordings are fed to the embedding function to compute their embedding vectors. In parallel, the labels are given to the triplet selection module so that it can list all the triplets that can be formed within the batch. The returned list contains indices of examples of the batch. The list of triplets and the embedding vectors are then passed to the triplet loss function (cf. Equation~\ref{eqn:tri}). Finally, the gradient of the loss with respect to the weights of the embedding function is computed and these weights are updated accordingly using the AdamW algorithm~\cite{loshchilov2017decoupled,kingma2014adam}.

  Diagram \textbf{B} describes the classification pipeline used to test the embedding functions  (cf.~Section~\ref{sec:embed_eval}). The pipeline takes as input an EEG recording, computes the embedding vector of this recording using the embedding function and passes it to the classification module. The classification module classifies the embedding vector either using a logistic regression classifier or a (one)-nearest-neighbor classifier. For details on the training of the classification module, please refer to Section~\ref{sec:training_scenarios}.}
\label{fig:train_test_diagrams}
\end{figure}

\subsubsection{Triplet loss} \label{sec:tri}
The original \textbf{triplet loss} \cite{schultz2004learning,weinberger2009distance}
is a loss function that takes as input triplets of embedding vectors that are an \emph{anchor} $\vectora=\rpz(\examplea)$,
a \emph{positive} example $\vectorp=\rpz(\examplep)$ and a \emph{negative} example $\vectorn=\rpz(\examplen)$.
Usually, the examples are labeled, with $\examplea$ and $\examplep$ coming from the same class but $\examplen$ from a different one.
In a \acrshort{bci} context, the labels can be, for example,
the name of the mental imagery task performed during the trial.
The triplet loss minimizes the distance between anchor and positive examples $\vdistap$
and maximizes the distance from anchor to the negative example $\vdistan$
such that:
\begin{align}\label{eqn:tri_ineq}
  \vdistap+\alpha < \vdistan
\end{align}
where $\alpha$ is a margin hyper-parameter.
The desired situation, for a given triplet $(\vectora,\vectorp,\vectorn)$, is visually represented in Figure~\ref{fig:space}\,\textbf{A}.
The triplet loss function is formulated as follows :
\begin{align}\begin{split} \label{eqn:tri}
    \loss^{\text{tri}}_\alpha(\tripletsindices, \vectorset)\eqdef
    \sum_{(\anchor, \positive, \negative)\in\tripletsindices}
    \maxzero{\vdistap-\vdistan+\alpha}
\end{split}\end{align}
where $\maxzero{\cdot}\eqdef \max(\cdot,0)$
and $\tripletsindices$ is a set of triplets of example indices.
To help with understanding, Equation~\ref{eqn:tri} can be reformulated as:
\begin{align}\begin{split} \label{eqn:tri2}
    \loss^{\text{tri}}_\alpha(\tripletsindices, \vectorset) =
    \sum_{\anchor=1}^N
    \loss^{\text{tri}}_\alpha \big( \{\anchor\}\times\indicesa{\oplus}\times\indicesa{\ominus}, \vectorset \big)
\end{split}\end{align}
with $\cdot\times\cdot$ being the Cartesian product, and, for all $\anchor$, $\indicesa{\oplus}$ is a set of indices of examples coming from the same class as the anchor $\anchor$, and $\indicesa{\ominus}$ are examples from other classes than $\anchor$.
Multiple different $\tripletsindices$ are valid and its creation is a matter of
triplet selection (cf. Section~\ref{sec:triplet_select}).\\
The training of an embedding function using a triplet loss is schematized in Figure~\ref{fig:train_test_diagrams}\,\textbf{A}.

\subsubsection{Ladder loss}
\begin{figure}[h!]
\centering
\includegraphics[width=.5\linewidth]{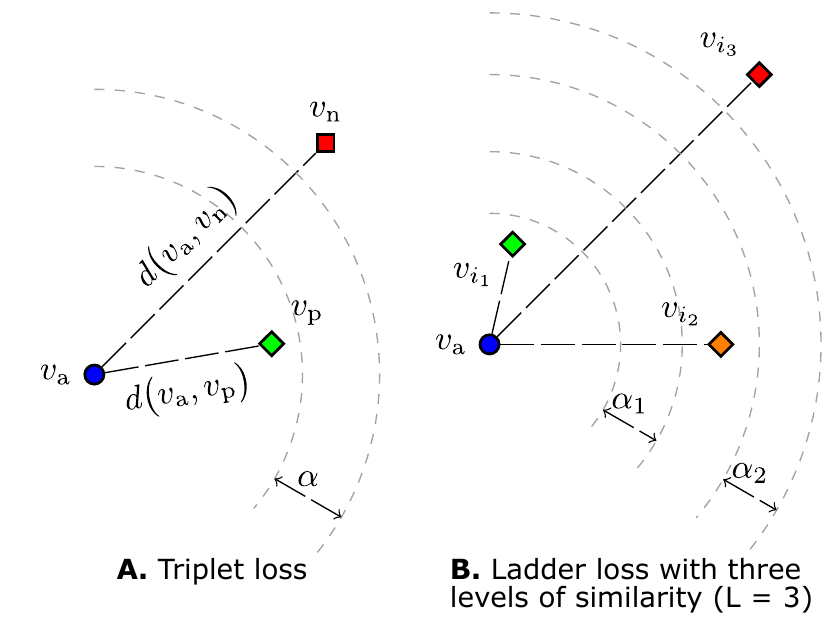}
\caption[Desired embedding spaces representations]{Desired embedding space when using \textbf{A} the original triplet loss
or \textbf{B} the ladder loss.}
\label{fig:space}
\end{figure}

The \textbf{ladder loss}~\cite{zhou2020ladder} is a recent extension of the triplet loss
where different "levels of similarity" to the anchor can be formulated ($l=1,2,\dots,L$).
The margins $\alpha_1, \alpha_2,\dots,\alpha_{L-1}$ are defined
for each consecutive level. The inequalities that the ladder loss tries to enforce are expressed by:
\begin{align}\begin{split}\label{eqn:lad_ineq}
  \vdista{i_1}+\alpha_1 &< \vdista{i_2} \\
  \vdista{i_2}+\alpha_2 &< \vdista{i_3} \\
  & \dots \\
  \vdista{i_{L-1}}+\alpha_{L-1} &< \vdista{i_L}
\end{split}\end{align}
Here the element $\anchor$ is an anchor for the elements $i_l\in\indicesa{l}$  considered per level $l$,
with $\indicesa{l}$ being a set of indices of examples with a similarity level of $l$ relatively to this anchor.
In Figure~\ref{fig:space}\,\textbf{B}, we can see an example of a well-arranged embedding space with three levels of similarity to the anchor.
To enforce this order on the distances,
the ladder loss is formulated as a weighted sum of triplet losses (one for each inequality):
\begin{align}\label{eqn:lad}
  \loss^{\text{lad}}(\vectorset)  &\eqdef \sum_{l=1}^{L-1} \beta_l
  \sum_{\anchor=1}^N \loss^{\text{tri}}_{\alpha_l} \big( \{\anchor\}\times\indicesa{l}\times\indicesa{l+1}, \vectorset \big)
\end{align}
with the constant parameters $\beta_1, \beta_2,\dots,\beta_{L-1}$ realizing the relative weighting of the triplet losses.
We recognize in Equation~\ref{eqn:lad} that the second formulation of the triplet loss (ie. Equation~\ref{eqn:tri2}) was used to express the triplet loss components.

\subsubsection{Product ladder loss} \label{sec:prod_lad_loss}
With the definition of the \textbf{product ladder loss}, we go beyond the original ladder loss as proposed by Zhou and colleagues~\cite{zhou2020ladder}.
It can be seen as a relaxation of the ladder loss
that allows for more flexibility when defining the examples' similarity levels to the anchor.
While the ladder loss requires a total order on the similarity levels,
the product ladder loss allows for non-total orders.
This means that with this novel loss it is possible to express the idea of having incomparable levels of similarity to an anchor.
We embrace this characteristic of the product ladder loss for the field of BCI, as
it allows to consider EEG recordings of multiple subjects performing multiple tasks as annotated by multiple independent labels simultaneously.
In such a case, some similarity comparisons can not be made (at least not without further assumption). Example: \textit{Are two recordings from the same subject performing different tasks more similar than recordings from two different subjects performing the same task?} We note that in this example no canonical total order on the similarities can be determined.

We suppose that examples have $K$ labels, all independent by default.
To evaluate the similarity level of two examples, we check for each label $k$
if the two examples have the same value on their $k$-th label or not.
This leads to a total of $2^K$ possible levels of similarity.
These levels are identified by a $K$ digits binary code $s_1 s_2\dots s_K\in\{0,1\}^K$ where,
for all $k$, $s_k=1$ if the compared examples share the same $k$-th label value and $s_k=0$ otherwise.
For any anchor $\anchor$ and any similarity level $s_1 s_2\dots s_K\in\{0,1\}^K$,
we can then define $\indicesa{s_1\dots s_K}$ as the set of indices of examples with this level of similarity to the anchor.
Finally, we have to configure our loss function by deciding how the similarity levels should be ordered and how this order should be enforced.
This task falls under the responsibility of the expert user and should be done by exploiting his domain knowledge.
To enforce an order, we define our loss as a weighted sum of triplet loss components
(similar to the ladder loss in equation~\ref{eqn:lad}).
The triplet loss components need, for each anchor $\anchor$, a positive set and a negative set which must be selected among the $\indicesa{s_1\dots s_K}$.

By \textbf{loss configuration}, we denote a list of triplet loss components,
i.e., a list of tuples that each contain a margin  $\alpha_i\in\mathbb{R}^+$, a weight $\beta_i\in\mathbb{R}^+$
and two similarity levels: one for the positive set $l^\oplus_i \in\{0,1\}^K$,
and one for the negative set $l^\ominus_i \in\{0,1\}^K$.
The order on similarity levels is not explicitly defined in the loss configuration
but follows from it.
Given a loss configuration, the product ladder loss is formulated as follows:
\begin{align}\begin{split} \label{eqn:prod_lad_loss}
    \loss^{\substack{\text{product} \\ \text{ladder loss}}}(\vectorset) &\eqdef \sum_{i} \beta_i
    \sum_{\anchor=1}^N \loss^{\text{tri}}_{\alpha_i}
    \Big( \{\anchor\}\times\indicesa{l^\oplus_i}\times\indicesa{l^\ominus_i}, \vectorset \Big)
\end{split}\end{align}
Later in this section, we will provide two concrete examples of loss configurations.

If no assumption is made on the labels, a product order on the similarity levels is proposed as the the default.
This order states that a level of similarity $s_1\dots s_K$ is higher than another level $t_1\dots t_K$ iff
$\forall 1\le k\le K, s_k\geq t_k$. 

We give here a broader explanation of why we came up with this loss function.
In most of the cases when learning an embedding with a triplet loss (face embedding, place embedding etc...),
the dataset contains a lot of different classes (it can go up to a few million classes).
The product ladder loss would allow to finely map the embedding space by showing the network an important diversity.
In motor-related BCI protocols however, the machine learning problem usually consists of classifying between two and four classes only.
Upon availability of metadata associated to the recordings,
it would be possible to define multiple sub-labels, and the number of possible combinations between them would increase exponentially with their number.
This gives us a way to artificially increase the number of different classes in our dataset.
Also, with this method, we suggest an embedding that can do more than just classify the main label: instead it paves the road for a future, multi-purpose embedding for EEG signals. In the following, we explore at which costs or benefits this versatility may come.

In this work, as a proof of concept, we only consider two labels:
the \subjectt label defining which subject was recorded during the trial
and the \targett label stating what imagery task was the subject performing during the trial.
This leads to four different levels of similarity:
The \subjectt label encodes the first component of the level's code
and the \targett the second component.
So supposing that two examples have a similarity level of $sc\in\{0,1\}^2$, then they have been recorded from the same subject
if $s=1$ and from different ones if $s=0$.
Similarly, the label $c$ will code for the imagery class.

\begin{figure}[h!]
  \centering
  \includegraphics[width=.5\linewidth]{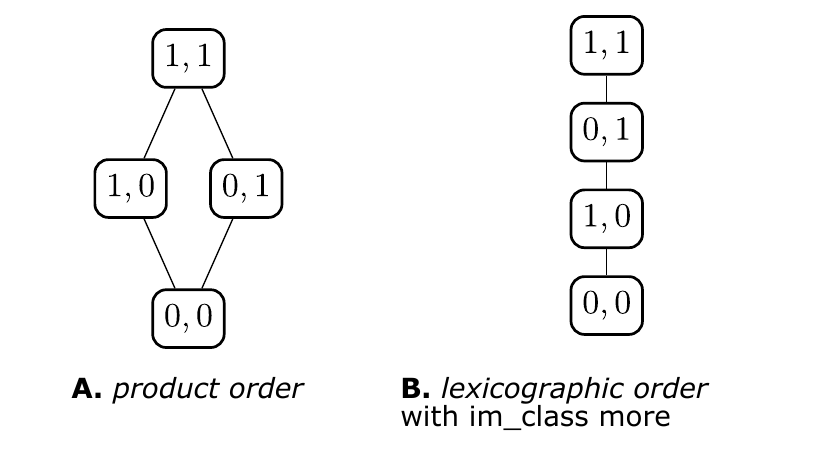}
  \caption[Hasse diagrams - distances ordering]{
    Hasse diagrams of the orders on similarity levels.
    Subfigure~\textbf{A} represents the diagram of the product order
    and subfigure~\textbf{B} the lexicographic order.
    In those diagrams, each node corresponds to a similarity level.
    If an edge goes upwards from a node $x$ to a node $y$,
    then the similarity level in node $x$ is lower than in node $y$.
    The edges that can be deduced by transitivity are not represented in these diagrams.
    This is why there is no direct edge, in both diagrams, between $0,0$ and $1,1$.
    We observe that in both diagrams, $1,1$ is always on top because it is always the highest similarity level
    and $0,0$ always at the bottom because it is the lowest.
  }
  \label{fig:configs_loss}
\end{figure}

For a given anchor $\anchor$, the sets of example indices of each of the levels are the following:
$\indicesa{11}$, $\indicesa{10}$, $\indicesa{01}$ and $\indicesa{00}$.
In this work, we only considered two types of orders on the similarity levels.
The first one is  \textit{product order}.
A visual representation of this order is presented in Figure~\ref{fig:configs_loss}\,\textbf{A}.
It results in the following loss:
\begin{align}\begin{split}\label{eqn:prod_lad_loss_prod}
  \loss^{\text{prod}}(\vectorset) = \sum_{\anchor=1}^N \Big(&
       \beta_1\loss^{\text{tri}}_{\alpha_1}(\{\anchor\}\times\indicesa{11}\times\indicesa{10}, \vectorset) \\
    +& \beta_2\loss^{\text{tri}}_{\alpha_2}(\{\anchor\}\times\indicesa{11}\times\indicesa{01}, \vectorset) \\
    +& \beta_3\loss^{\text{tri}}_{\alpha_3}(\{\anchor\}\times\indicesa{10}\times\indicesa{00}, \vectorset) \\
    +& \beta_4\loss^{\text{tri}}_{\alpha_4}(\{\anchor\}\times\indicesa{01}\times\indicesa{00}, \vectorset) \Big)
\end{split}\end{align}

The second one is a \textit{lexicographic order} where the \targett component
is considered more important than the \subjectt one.
A visual representation of this order is presented in Figure~\ref{fig:configs_loss}\,\textbf{B}.
This order is total and results in the following loss:
\begin{align}\begin{split}\label{eqn:prod_lad_loss_lexi}
  \loss^{\text{lexi}}(\vectorset) =  \sum_{\anchor=1}^N \Big(&
        \beta_1\loss^{\text{tri}}_{\alpha_1}(\{\anchor\}\times\indicesa{11}\times\indicesa{01}, \vectorset) \\
    +& \beta_2\loss^{\text{tri}}_{\alpha_2}(\{\anchor\}\times\indicesa{01}\times\indicesa{10}, \vectorset) \\
    +& \beta_3\loss^{\text{tri}}_{\alpha_3}(\{\anchor\}\times\indicesa{10}\times\indicesa{00}, \vectorset) \Big)
\end{split}\end{align}

\subsubsection{Triplet selection} \label{sec:triplet_select}

\emph{Hard triplets} are triplets which challenge the minimization of the triplet loss.
Given an anchor $\anchor$,
we want to find positive examples $\positive$ maximizing $\vdistap$
and negative examples $\negative$ minimizing $\vdistan$.
These are triplets that the network is not successfully placing in the embedding space.
The selection of these \emph{hard triplets} is a central point of the triplet loss approach
, and is also called triplet mining.
It is crucial because the number of potential triplets
grows cubicly with the dataset size.
If only \emph{easy} triplets are presented to the network,
then the training might take very long.
However, determining the triplets by computing an $argmax$ and an $argmin$ on a full dataset is computationally unfeasible and inefficient.
Instead, mining strategies have been proposed to determine triplets.

Triplet mining can be done offline or online. In an offline strategy, the current network checkpoint is used to compute the representations of the dataset's elements.  Then, specifically suitable, i.e. hard, triplets can be selected according to various criteria.

During online selection, only the $n$ training examples contained in the current  batch would be considered to form possible triplets. Again, suitable triplets would be selected among those only.

Common selection criteria are to pick the top $k$ triplets of those which do not respect a desired condition.

Clearly, the online selection has a computational advantage as a smaller number of training examples need to be mapped into the current embedding, which is the most computationally expensive part. On the other hand, online selection might reduce the diversity among the triplets because they all stem from
the same $n$ examples contained in a batch. Please note that, depending on the dataset, there might not be a lot of positive matches contained in the current batch, specifically if the number of classes is high.
So, to maximise their number, the $n$ batch elements can be sampled from a subset of the dataset.
For example, all examples can come from a subset of the labels
so that the probability that two batch elements share the same label is increased.

Also, we can note that the $\maxzero{\cdot}$ function in the loss
realizes a triplet selection by itself, as only those triplets $(\anchor, \positive, \negative)$
violating the condition $\vdistap+\alpha\le\vdistan$ have an influence upon the training.

Mining triplets for a ladder loss or a product ladder loss is not very different
than for the triplet loss.
Examples can simply be mined using the methods described above for each of the components independently.
In this work, we have opted for an online triplet mining strategy
where batch examples are sampled from a subset of the dataset.
Our only addition comes from the fact that each example has multiple labels in our case.
We define the subsets such that they restrict the number of possible classes per label.
Furthermore, in every batch we balance the number of examples having each combination of labels.
We do not perform any selection of hard triplet because
we did not deem it necessary given the small size of our dataset.

\subsection{Metrics}
We will use, throughout this paper,  different classification metrics.
The first one is the classification \textbf{accuracy}.
It is computed as the number of test examples that are correctly classified by the model
divided by the total number of test examples.
The second metric, only mentioned in the confusion matrices, is the \textbf{recall} score.
We compute a recall score for every class involved in a classification task
and it corresponds to number of examples that were correctly attributed to this class by the classifier (i.e. the true positives)
divided by the number of examples that are truly belonging to that class (i.e. the true positives plus the false negatives).
Finally, we will report in the confusion matrices the \textbf{precision} score.
We also compute a precision score for every class involved in a classification task
and it corresponds to number of examples that were correctly attributed to this class by the classifier (i.e. the true positives)
divided by the total number of examples that were attributed to that class by the classifier (i.e. the true positives plus the false positives).

\subsection{Evaluation of Embeddings} \label{sec:embed_eval}

As embeddings can not be reduced to a single purpose,
one metric to assess them may not be sufficient. An obvious metric is provided by utilizing an embedding as a preprocessing for a classification task and reporting the classification accuracy, which is in line with most previous approaches for representation learning and metric learning applied to EEG/BCI data.

For this purpose, embeddings will be trained using a triplet- or ladder loss, without incorporating any feedback from the classifier.
Once an embedding function is fully trained,
the embedding vectors of the classifier's train and test sets are computed.
The classifier is then trained on the embedded vectors,
and its accuracy on the test set is reported as the score of this particular combination of an embedder plus classifier.
In our experiments, we will use a logistic regression algorithm to classify the embeddings.
We use the Scikit-learn implementation~\cite{scikit-learn} with default parameters,
ie. with an L2 penalty, a regularization strength of $1.0$ and  100 iterations.
But we will also compare it to a very simplistic one-nearest-neighbor classifier. The reasoning behind this choice is that this comparison may provide an intuition, if the embedding or rather the classification algorithm contributes to the observed overall performance.

The pipeline used to classify the embedding vectors during the test phase is schematized in Figure~\ref{fig:train_test_diagrams}\,\textbf{B}.

Nevertheless, we will keep in mind that this evaluation approach using a classification task only offers a partial view of the value of an embedding
and that embeddings may provide a suitable representation also for other machine learning tasks.

\subsection{Training scenarios}\label{sec:training_scenarios}

The authors of the \acrlong{hgd} provided a train/test split of the data, with approx.~880 train trials and exactly 160 test trials per subject. We adopted this split to stay consistent with other studies using the dataset.

We will compare different training scenarios.
They have been defined to cover cases where (1) no time at all can be allocated for a calibration prior to running a online experiment, (2) limited time is available to record some calibration data but not sufficient time to train a deep neural network from scratch, and (3) where there is enough time to do both.
We will analyze, for each of them separately, if it is interesting to use embeddings or not. The scenarios are described below:

\paragraph*{Within-subject scenario:}
For each experiment (\acrshort{fbcsp}, \acrshort{cnn} classifier or embedder),
one model is trained for each subject on the predefined train part of this subject's data
and tested on the predefined test part.
In the embedding case, the embedding classifiers must also be trained (cf. Section~\ref{sec:embed_eval}).
So once the training of an embedding function is finished, the embedding vectors of the train and test part of the corresponding subject's data are computed.
The embedding classifier is trained on the embeddings of the train part and tested on the embeddings of the test part.
For a given embedding function, multiple scores are available
because multiple classifiers can be trained on top of the same embedding.
For each model, the average accuracy across all subjects is reported along with the individual scores.

\paragraph*{Cross-subject scenario:}
In this case, the model is trained on the data coming from all subjects but one
and tested on the test part of the held-out subject's data.
So for each experiment (\acrshort{cnn} classifier or embedders),
we train as many models as there are subjects in the dataset (like in the within-subject case).
When the trained model is an embedder, we additionally have to fit a classifier.
Here we will distinguish two different scenarios:
\begin{itemize}
  \item \textbf{Complete \acrshort{loso}} (for \acrlong{loso}): In this scenario, we train the embedding classifier on the data of the same thirteen subjects, which were used to learn the embedding function. We test the classifier on the embedded test data points of the held-out subject.
  \item \textbf{Partial \acrshort{loso}}: In this scenario, we assume that for a novel subject some data is available from a calibration recording. Thus, we
train the classifier on the embedded training data points of the held-out subject,
  and we test it on the embedded test data points of the same held-out subject.
\end{itemize}

Please note that any test score of any cross-subject model for a given held-out subject is computed on the same data points as those used to compute the test score
of any within-subject model for the same subject.
This will allow us to compare the within-subject models with the cross-subject ones.\\

In this paper, we report the results obtained by 21 different models.
Each of them was trained once with each subject as either held-out subject (cross-subject transfer learning scenarios) or the only subject (within-subject scenarios).
In a within-subject scenario, we trained \acrshort{fbcsp} and EEGNet in a simple classification configuration. In addition, EEGNet is trained as an embedding function using the original triplet loss.
The product ladder loss would be useless in this case because one subject only is present in the train set.
In a cross-subject scenario, we trained again EEGNet as a classifier and four EEGNet models as embedding functions, each with a different loss configuration.
The first configuration, named \textbf{(a) \desca}, is the original triplet loss where the embedding model was trained to only separate the motion classes executed during the trial.
The three other configurations are product ladder losses that consider two labels: the imagery class (\targett) and the subject (\subjectt).
The second configuration, named \textbf{(b) \descb}, supposes a lexicographic order on the examples with
the \targett label being more important than the \subjectt label (cf. Section~\ref{sec:prod_lad_loss}).
We weight the triplet loss components using $\beta_1=\beta_3=1$ and $\beta_2=3$ in equation~\ref{eqn:prod_lad_loss_lexi}, such that the importance of separating the motion classes is empathized by a factor of three.
The third configuration, named \textbf{(c) \descc}, is also a lexicographic order but with equal weight on the three components.
The fourth configuration, named \textbf{(d) \descd}, is a product order with equal weight on all four components.
With these four loss configurations, we implemented a gradual increase
of the importance given to the separation between the subjects.
For each embedding, we report the accuracy score obtained by logistic regression and one-nearest-neighbor classifiers
and, for all the cross-subject embeddings, these accuracies are computed both
in partial \acrshort{loso} and complete \acrshort{loso} manner.
\newcommand{\descEmbedding}[1]{embedding #1
  (with four classification pipelines: with logistic regression or nearest neighbor classification
  and in complete or partial \acrshort{loso})}
The exhaustive list of the 21 classification pipelines we compare is given below.
The first four are the within-subject pipelines:
\begin{itemize}
  \item \acrshort{fbcsp}
  \item EEGNet
  \item embedding~ ($\times2$)
\end{itemize}
For cross-subject analyses, 17 pipelines were utilized:
\begin{itemize}
  \item EEGNet
  \item embedding (a) \desca~ ($\times4$)
  \item embedding (b) \descb~ ($\times4$)
  \item embedding (c) \descc~ ($\times4$)
  \item embedding (d) \descd~ ($\times4$)
\end{itemize}
Each within-subject embedding model leads to two classification pipelines, one with
a logistic regression classifier and one with a nearest neighbor classifier.
The cross-subject embedding models each result in four classification pipelines,
depending again on the classifier used on top of the embedding (logistic regression or nearest neighbor) and on the classifier's training scenario (complete or partial \acrshort{loso}).

\subsection{Statistical analysis}\label{sec:stats}

A model is trained fourteen times, each time with a different test subject.
As the performance scores can not be expected to follow a normal distribution, comparisons between two models will be based on the rankings of the test subjects.
We statistically test this with
Wilcoxon signed-rank tests~\cite{wilcoxon1992individual}.
Whenever a statistical significance is reported,
we refer to a p-value lower than $0.05$.
To account for multiple tests conducted,
we apply a  Holm–Bonferroni correction~\cite{holm1979simple}.

\section{Results} \label{sec:results}


We will first present the results obtained with existing baseline methods, before we investigate if the use of information about the subjects is helpful for discriminating the four imagery classes. Then, comparisons between embeddings and baseline methods will be performed for different scenarios, before embedding spaces are analyzed with respect to potentially learned structures.
In all the accuracy comparisons, as the four classes are balanced, the chance level is at $25\%$.

\subsection{Performances of Baseline Methods}
\label{result:baseline}

Among the previously published methods, \acrshort{fbcsp} is a relatively fast method which could be preferred over EEGNet when calibration time is key. Applied as a baseline method in within-subject scenarios,  it reaches an average accuracy across subjects of $82.8\%$. While we found that it does not perform competitively for LOSO or partial LOSO scenarios, \acrshort{fbcsp} could still be used in partial LOSO scenarios in a within-subject mode on the training data of the held-out subject only, as its fast training would not introduce a substantial delay between recording the training data and applying it during the online application.
As a baseline method for neural network models, for which the training takes substantially longer and may require the fine-tuning of several hyper-parameters, we report the classification scores of EEGNet. On average over all subjects, EEGNet reaches accuracies of $86.8\%$  for the within-subject scenario and $67.7\%$ across-subjects in a full LOSO scenario.

\begin{figure}[h!]
  \centering
  \includegraphics[scale=.5]{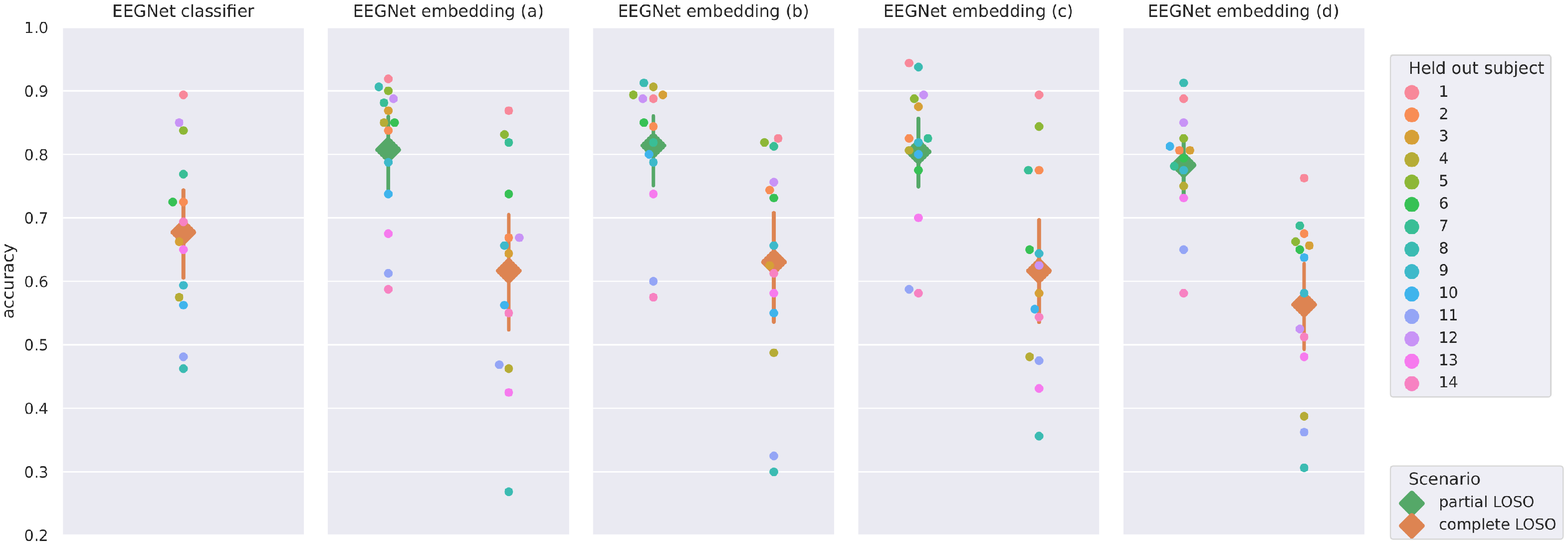}
  \caption[Cross-subject classification performance (lexicographic classifiers)]
  {Classification performance for \textbf{cross-subject} training.
  Each grey box contains the results of a different model.
  The boxes of the embedding models each contain two pipelines corresponding to the complete and partial LOSO scenarios.
  Large markers represent the average across subjects and bars denote the standard errors,
  the orange ones are for partial LOSO models and the green ones for complete LOSO.
  Small dots represent the individual scores obtained with each of the held out subjects.
  The embedding vectors are classified using logistic regression classifiers.
  For the results with nearest-neighbor classification, see the Supplementary Figure~\ref{fig:acc_cross-subj_1nn}
  }
  \label{fig:acc_cross-subj_logistic}
\end{figure}

\subsection{Comparison between embeddings}
\result{Learning to separate the subjects may help separating the imagery classes. \label{result:embed_comp}}
In the context of cross-subject embeddings, we analyzed the performances of the four product ladder loss configurations. Please note that in loss configuration (a) only the motion classes have to be separated by the embedding, while in loss (d) both the classes and the subjects are supposed to be separated (with equal importance on both) by the embedding training. With configurations (b) and (c), we realize a gradual increase in the importance of separating the subjects
(cf. Section~\ref{sec:training_scenarios}).
The results of the different models are presented in
Figure~\ref{fig:acc_cross-subj_logistic} for a logistic regression classifier and in Supplementary Figure~\ref{fig:acc_cross-subj_1nn} for the one-nearest-neighbour classifier.
Green markers denote the average accuracies across subjects computed in partial LOSO
while orange markers denote complete LOSO.
We observe that the embedding model with best score on average is model (b),
regardless of the classifier used, and we observed this both for the complete LOSO and partial LOSO scenarios. Model (b) on average showed an improvement of $1.39\%$ and $0.67\%$ over model (a) for the complete LOSO and partial LOSO scenarios respectively (both with logistic regression classifiers). However, as this small difference did not reach significance based on a two-sided Wilcoxon signed-rank tests (p-value 0.05), the limited number of subjects in this dataset does not allow to make a final judgement on whether learning to separate the subjects is helpful also for separating the four imagery classes.

Nevertheless, we will use loss configuration (b) in the following to compare to the different baseline methods.

\subsection{Comparison of the embeddings to the state-of-the-art}
\result{On complete LOSO, our embeddings do not lead to a better generalization than vanilla classification.}
In cross-subject experiments, we compare the baseline EEGNet classifier to embeddings classified in complete LOSO scenarios. Please note that the embedders and their classifiers are trained on the same data in these scenarios.
The results are visualized in Figure~\ref{fig:acc_cross-subj_logistic},
where the leftmost subplot depicts the accuracies obtained by the baseline EEGNet. The best embedding function (b)
has an average accuracy of $63.0\%$, which is  $4.7\%$ below that of the baseline EEGNet model.
While this difference is not statistically significant (Wilcoxon signed-rank test, p>0.05), we found no evidence for an advantage when using an embedding function in the full-LOSO scenario of a cross-subject problem.

\result{For partial LOSO scenarios, embeddings surpass the baseline EEGNet.}
In this result, we assume a cross-subject experiment, where limited calibration data of the test subject has been collected. As the time limitation is too prohibitive for the baseline EEGNet classifier to be re-trained, it is restricted to a complete LOSO scenario. An embedding pre-trained on all but one subjects, however, could be equipped with a (logistic regression) classifier that can be trained quickly on the novel data points (partial LOSO).
In this situation, the embedding functions are exactly the same as in complete LOSO (trained on all but one subject)
but we suppose that we have access to some data of the test subject to fit the embedding classifier.
This scenario can occur, for example, if we allow a short offline training before
the online application starts.
We assume that in a partial LOSO scenario it would be difficult to train or fine-tune a deep neural network due to the substantially longer training time (minutes instead of sub-second), and as typically the monitoring of the training progress is required.
Fitting a logistic regression classifier or a full \acrshort{fbcsp} pipeline however is feasible as it is quicker and as their training procedures can easily be automatized.

Again the relevant accuracies for EEGNet in complete LOSO and of embeddings combined with logistic regression are  reported in Figure~\ref{fig:acc_cross-subj_logistic}.
The best partial LOSO embedding pipeline (b) obtains an average accuracy of $81.4\%$.
This is $13.8\%$ higher than the cross-subject baseline performance of EEGNet
and this difference is statistically significant (Wilcoxon signed-rank test, p<0.05).
We observe a huge jump in accuracy
for an insignificant overrun computing cost. This kind of "low-cost fine-tuning" is not feasible in the case of the baseline EEGNet classification model.

\begin{figure}[h!]
  \centering
  \includegraphics[scale=.5]{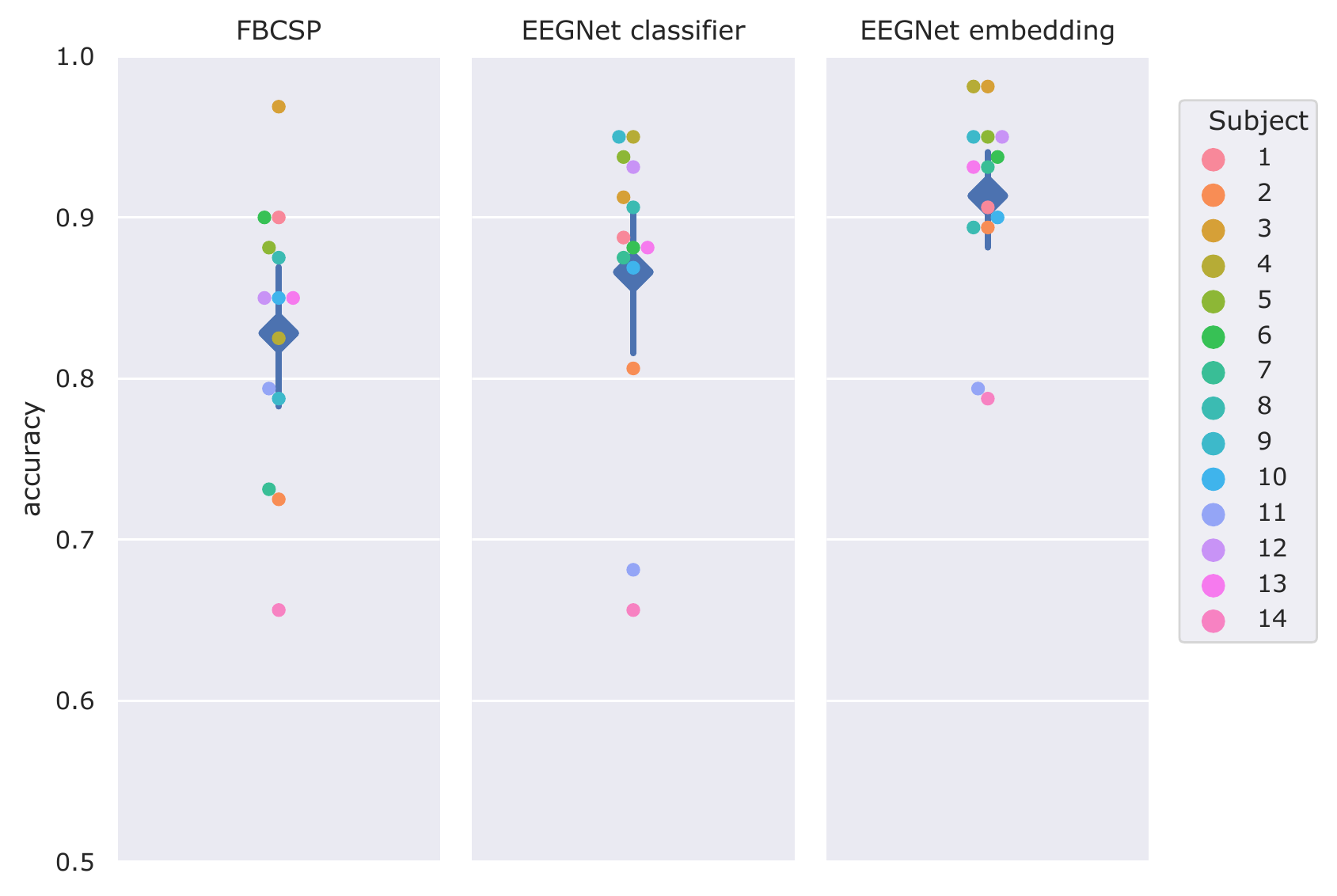}
  \caption[Within-subject classification performance (lexicographic classifiers)]
  {Classification performance on \textbf{within-subject} training.
  Each grey box contains the results of a different model.
  Large markers represent the average across subjects and bars correspond to standard errors.
  Small dots represent the individual scores obtained for each of the subjects.
  The embedding vectors are classified by logistic regression classifiers.
  For the results with nearest-neighbor classification, see the Supplementary Figure~\ref{fig:acc_singl-subj_1nn}. }
  \label{fig:acc_singl-subj_logistic}
\end{figure}

\result{In partial LOSO scenarios, embeddings approximate the score of a within-subject \acrshort{fbcsp} baseline.
\label{result:partial_loso_fbcsp}}
When trials of the test subject are available, training a \acrshort{fbcsp} pipeline from scratch on this data
(ie. within-subject training) requires relatively little computing effort.
Thus within-subject performance of \acrshort{fbcsp} is compared in the following to the performance of embeddings equipped with logistic regression trained in a partial LOSO scenario.

The scores of the embedding models are provided by green markers in Figure~\ref{fig:acc_cross-subj_logistic}, while the \acrshort{fbcsp} baseline performances are depicted in Figure~\ref{fig:acc_singl-subj_logistic}.
Comparing the best embedding model (b) to \acrshort{fbcsp}, we observe a $1.4\%$ difference in favor of \acrshort{fbcsp}, which is not statistically significant (two-sided Wilcoxon signed-rank test, p>0.05).
Features extracted by the CSP in the \acrshort{fbcsp} pipeline
are classified by a logistic regression algorithm (cf. Section~\ref{sec:fbcsp}),
as are the embedded vectors in the proposed model.
And, while the embedding function can always be improved by enlarging the dataset for pre-training the embedding, \acrshort{fbcsp} can not be enhanced easily.
The question if the embedding classifier might be trainable with even less data
from the test subject than \acrshort{fbcsp} will be addressed in Result~\ref{result:few_samples_fbcsp}.

\begin{figure}[h!]
  \centering
  \includegraphics[height=7cm]{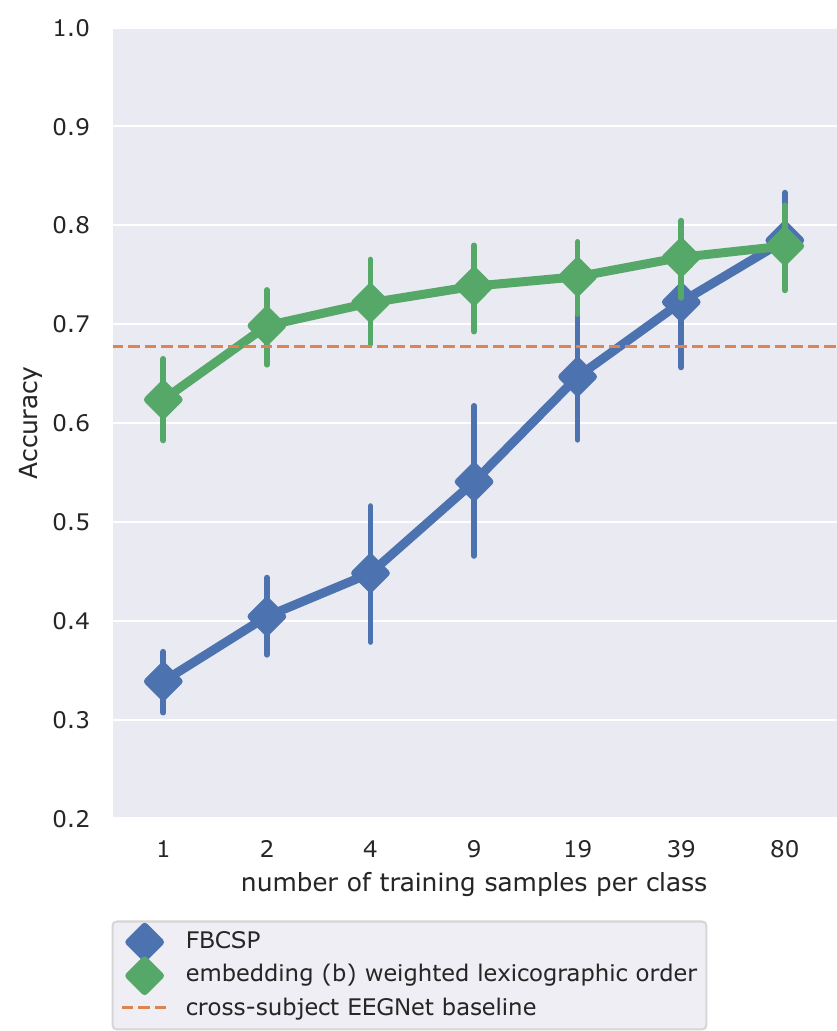}
  \caption[Fine-tuning on few samples]{
    Development of the classification accuracy for an increasing number of training samples per class to calibrate
    either the FBCSP baseline method (within-subject scenario, denoted in blue) or embedding configuration (b) with a logistic regression classifier
    in a partial LOSO scenario (denoted in green).
    Markers denote the averages over fourteen subjects, while the vertical bars represent the standard errors.
    The training samples have been provided to the classifiers in chronological orders as recorded during the session of the test subject. The limitation to 80 samples per class is given by the subject with the lowest trial count contained in the dataset.
    The performance of the cross-subject EEGNet baseline model is marked by the orange dotted line.
    The results obtained by the other partial LOSO models can be found in Supplementary Figure~\ref{fig:few_samples_all}.
  }
  \label{fig:few_samples}
\end{figure}

\begin{figure}[h!]
  \centering
  \includegraphics[width=\linewidth]{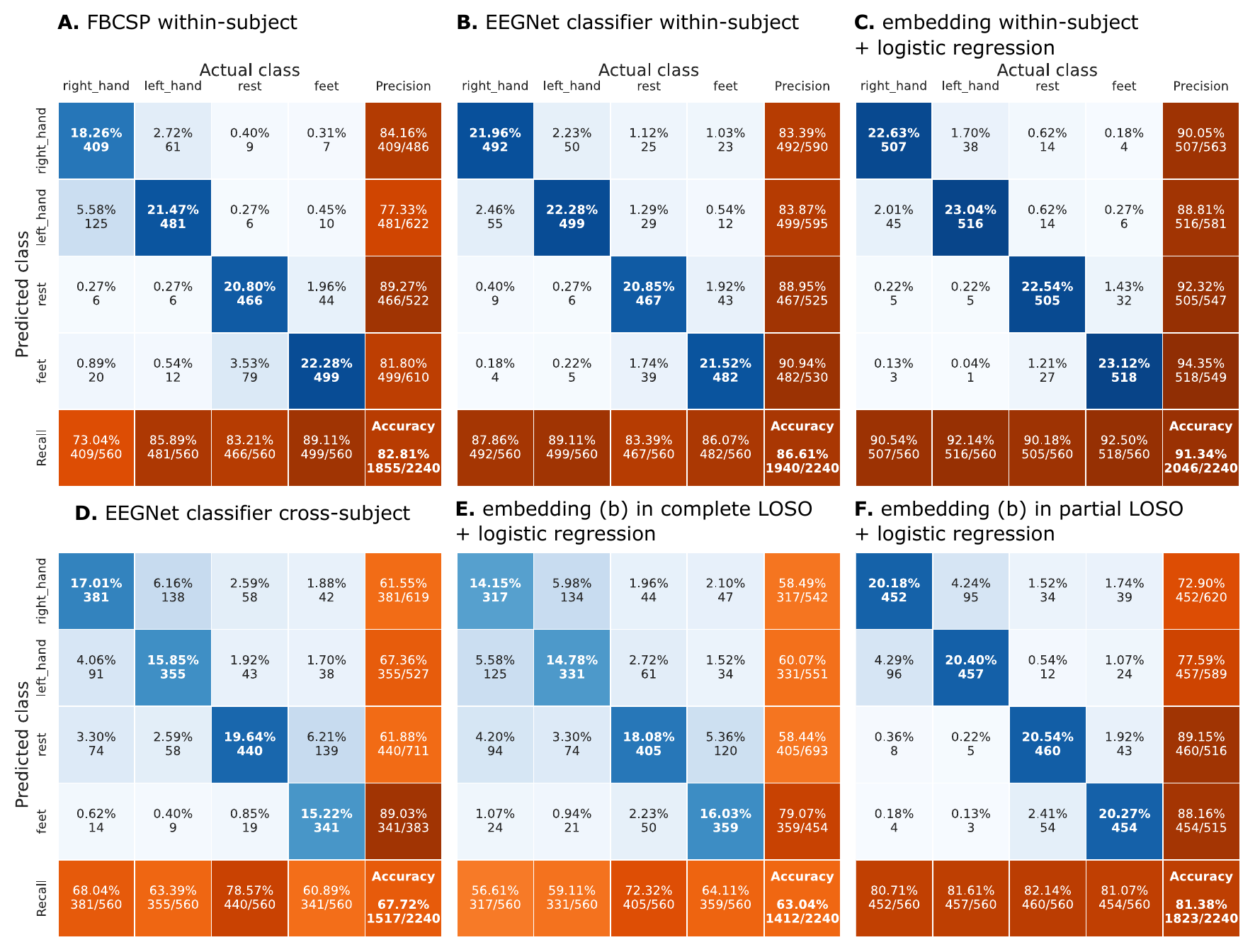}
  \caption[Confusion matrices]{Confusion matrices of six different classification pipelines.
    In each matrix, the cell at the bottom right is the overall accuracy of the model,
    the other cells of the last column contain the precision of the model for each of the classes
    and the cells in the last row contain the recall scores for each class.\label{fig:confmat_main}}
\end{figure}

\result{Few trials of a new subject are sufficient for an embedding classifier to surpass the cross-subject baseline EEGNet.} \label{result:few_samples_cross}
Over subjects, the amount of training data varies between 80 and 224 trials per motor task. In the partial LOSO scenarios, the data of the left-out subject is not used to train the embeddings, but is utilized to train a classifier on top of the embedding function learned on the remaining subjects. The amount of test trials per subject and per motor task is always 40. So far, results of the partial LOSO models presented in Figure~\ref{fig:acc_cross-subj_logistic}
and in the corresponding confusion matrices (Figure~\ref{fig:confmat_main})
were obtained using the maximum of the available training data to calibrate the classifier on top of the embedding functions.
\\
As embedding vectors only have eight dimensions and as the logistic regression and the one nearest-neighbour classifier both have a relatively low capacity, we investigated how a reduction of calibration data points impacts upon the performance obtained on the 40 unseen test data points. The results are visualized in Figure~\ref{fig:few_samples}.
\\
We observe that the classifiers already perform well with very few subject-specific calibration data points (green curve), and one representative per class is sufficient to score way above chance level.
With two training samples per class, this model on average can beat the cross-subject EEGNet baseline, whose performance is indicated by the orange doted line. This baseline model was used in a complete LOSO scenario.
This amount of calibration data corresponds to 8 trials of four seconds duration, which can be recorded in 56 seconds if inter-trial pauses of 4 seconds are taken into account. As the training of a logistic regression classifier takes less than a second on this data, the total calibration time (including data acquisition) required to beat the cross-subject EEGNet baseline is less than a minute.
Our observation that the embedding classifier can be trained with very little subject-specific training data indicates that the embedding vectors probably form a structure, which is highly informative for the classification task.

\result{The partial LOSO model requires less calibration samples than the FBCSP baseline} \label{result:few_samples_fbcsp}
In the same way as described above, we recall that the results of within-subject FBCSP scenario presented in Figure~\ref{fig:acc_singl-subj_logistic} and in the confusion matrices were obtained using the whole train set of the subjects as calibration data.
However, one can expect that the FBCSP also suffers from smaller amounts of calibration data. Thus we reduced the calibration data available. The resulting test performance is provided by the blue curve in  Figure~\ref{fig:few_samples}.
We observe that for small numbers of samples, the embeddings with calibrated logistic regression classifier (the partial LOSO scenario) outperforms FBCSP by up to 30\,\% accuracy.
The FBCSP baseline needs at least around 80 training samples per class to reach the performance of the embedding model, which corresponds to a calibration time of approximately 43 minutes. If the calibration phase is expected to be shorter, our results propose to prefer the embedding function over FBCSP.

\result{Embedding is beneficial even in the within-subject scenario.}
\label{result:within_subject}
Finally, in case one has both access to some of the test subjects' data
and enough time to train any desired model, it is possible to compare any baseline model with embedding models, when both are trained in a pure within-subject scenario. The scores of these within-subject models are shown in Figure~\ref{fig:acc_singl-subj_logistic}.
The first obvious model choice is the within-subject classification using the EEGNet baseline, which surpasses the \acrshort{fbcsp} baseline by an insignificant difference of 4\,\% accuracy (two-sided Wilcoxon signed-rank test, p>0.05).
The within-subject embedding, trained on data of one subject only, performs best among all our models.
In particular, it outperforms the FBCSP model by a margin of 8.5\,\% accuracy (not tested for statistical significance), and it significantly surpasses the baseline within-subject EEGNet by 4.7\,\% on average (Wilcoxon signed-rank test, p<0.05).
The latter comparison is specifically interesting as the two models are very similar to each other regarding the use of a logistic regression for the final classification step and as both are trained on exactly the same data.
Please note, that this within-subject embedding model at least keeps up with the results reported by Schirrmeister et al., who published the dataset and proposed shallow and deep convolutional architectures for this data~\cite{schirrmeister2017deep} despite we did not provide our models with information of the high gamma band (see details in Section~\ref{sec:dataset}).


\result{Embedding models and baseline models suffer from the same difficulties. \label{result:confmat}}
The embedding models and the baseline models show similar difficulties, specifically to discriminate left-hand and right-hand trials as well as resting state and feet trials, as reported in the confusion matrices of six different classification pipelines in
Figure~\ref{fig:confmat_main}.
The confusion matrices of other models and configurations
are available in the Supplementary Figure~\ref{fig:confmat_embeddings_appendix}.
We observed that in complete LOSO scenarios (see subplots \textbf{D} and \textbf{E} in Figure~\ref{fig:confmat_main}), additionally a stronger confusion between resting state and both hand classes plays a role, with slightly smaller problems to discriminate between feet and both hands. Resolving these confusions seems to be one of the main improvement introduced when switching from complete LOSO to partial LOSO (cp.~subplots \textbf{E} and \textbf{F} in Figure~\ref{fig:confmat_main}), which is enabled by the embeddings.

\subsection{Visualization of embedding spaces}

\embedFigure{tsne_all}{2}{\acrshort{tsne}}{
  The colors encode the subjects
  and the shapes encode the task classes.\label{fig:tsne_all}
}
\embedFigure{tsne_im-class}{2}{\acrshort{tsne}}{
  The colors encode the task classes.
  \label{fig:tsne_im-class}
}

\result{Embedding spaces reveal structures.}
To introspect the embedding spaces, we project the 8-dimensional vectors using \acrshort{tsne} \cite{van2008visualizing} and \acrshort{umap} \cite{mcinnes2018umap} into two-dimensional maps. Both projection methods try to maintain a balanced local versus global distance relationship of the vectors between the two spaces.
We can subjectively observe the quality of the embedding spaces in
Figures~\ref{fig:tsne_all} and~\ref{fig:tsne_im-class} for the \acrshort{tsne} projections.
The \acrshort{umap} projections are available in the additional materials as
Supplementary Figures~\ref{fig:umap_all} and~\ref{fig:umap_im-class}. Overall we observed that both, the \acrshort{tsne} and \acrshort{umap} visualizations are rather consistent.

The loss configurations (c) and (d) depicted in Figure~\ref{fig:tsne_all} use color to code for subjects. The 2D projections show a clear separation of subjects into different disjoint clusters. This separation is not as clear for configuration (b) and not present at all in (a).
This observation is consistent with our expectations as the four embeddings were designed with different emphasis on the subject condition.
In configurations (c) and (d) we observe that the exemplarily added test subject number 2 forms a separate cluster, but this was not the case for all test subjects.
In Figure~\ref{fig:tsne_im-class} the same embeddings are depicted, but here the colors code for the task classes.
In the subplots of configurations (a) and (b), four main clusters emerge, which correspond to the four task classes.
This is consistent with the fact that their underlying embeddings were trained with stronger weight on separating the classes.
With configurations (c) and (d), we observe that per subject the classes form sub-clusters. In case no clear sub-clusters could be identified, this mainly concerned the datapoints of right and left hand classes (blue and orange) or the rest and feet classes (green and red).
This suggests that these two pairs of classes are harder to distinguish than other pairs, which is consistent with our interpretation of the confusion matrices
(cf. Result~\ref{result:confmat}).

\section{Discussion} \label{sec:discussion}

We have shown how a neural network can embed EEG signals
when the embedding is trained using a metric distance defined by the product ladder loss.
We have also shown how the specific configuration of this loss influences
the classification performance and shapes structures that emerge in the embedding space.
In the following, we will discuss hyperparameter choices made and selected results.

\discussion{Embedding dimensionality choice.}
In all our experiments, the embedding spaces have eight dimensions, which is relatively small compared to the standard metric learning practice in other fields.
We decided to commit to this value based on a preliminary experiment, for which
we compared nine different embedding dimensions evenly spaced on a logarithmic scale
(1, 2, 4, 8, 16, 32, 64, 128 and 256 dimensions).
We compared them on different loss configurations and embedding scenarios.
For all combinations, we observed a systematic increase in classification accuracy until eight dimensions were reached, and a saturation of the performance for higher dimensions.
We are aware that the dimensionality parameter might have been overfitted on our test data, but  the monotonicity of the increase in accuracy with the dimensionality observed on all the different configurations until eight dimensions are strong indications that the choice of eight dimensions may be a generally suitable one for this dataset.
In future experiments, it nevertheless might be interesting to explore the influence of the embedding's dimensionality, for example if a more expressive network is trained to learn an embedding
or if a larger or richer dataset with additional metadata can be used.

\discussion{Choice of the loss function.}
In this study, we use a variation of the triplet loss to learn embeddings
with an Euclidean distance in the embedding spaces.
Extended future studies will show if metric learning could lead to better cross-subject transfers
and possibly also to better cross-task transfers
than classification-based representation learning can.
With a latent representation learned using a classification loss,
the only goal is to make the features separable with respect to the specific classification task.
On the contrary, embeddings learned using a metric-learning loss
are optimized to be discriminative \cite{wen2016discriminative}.
Musgrave and colleagues~\cite{musgrave2020metric}
showed that the successive generations of metric learning losses,
all supposed to outperform their predecessors according to their authors,
only performed marginally better than classic methods.
While considering their study, we chose to use and modify a triplet loss over using other metric learning loss functions
because the triplet loss is generally well known in the machine learning community.
The design of the ladder loss allows to exploit prior knowledge of the data.
We simply extended it to satisfy the constraints on our dataset.

\discussion{Superiority of embeddings in within-subject scenarios.}
In Result~\ref{result:within_subject}, we observed that
 the triplet loss obtains a better score than the classification loss in the within-subject-scenario.
This directly supports the conclusion of Alwasiti et al.~that metric learning for the purpose of classifying EEG signals allowed them to train a deep neural network with a relatively small number of training samples (\textasciitilde 120 EEG trials)~\cite{alwasiti2020motor}.
We found their work very inspiring, but would have been interested to directly see a comparison of their approach
with the straightforward classification loss.

In our within-subject scenario analysis, we conducted such a comparison.
Even if we have more training samples in the \acrlong{hgd} (between 320 and 896 training samples per subject),
it remains relatively small compared to training data sets used in other machine learning fields.
Nevertheless, we observed a strict advantage for the metric-based within-subject loss with few training samples.
\\
In the literature, the architectures trained by triplet loss or other metric-based losses are sometimes referred to as
\textit{Siamese networks} or \textit{Triplet networks}~\cite{chicco2021siamese}.
and it typically is reported that Siamese networks
can obtain good results  compared to regular classifiers  when trained on small datasets~\cite{bromley1993signature}.
This is due to the fact that they are trained to find the differences and similarities between data points
instead of just trying to classify them.
In a way, the feedback provided to the network by a metric-based loss is more informative
than from classification losses.
Siamese networks are used up to the extreme of one-shot learning~\cite{fei2006one}
where only one or few examples per class are available.

\discussion{Use of data augmentation.}
As metric learning with triplets allows to exploit a small dataset well, we decided not to use data augmentation techniques in addition. However, this could be a relevant next step for algorithm development and benchmarking~\cite{castano2019post} but also for translating BCI methods from labs to bedsides, as domain-specific augmentations could lead to embeddings that are invariant to artifacts, noise or may even be useful when working with a varying number of electrodes.

\discussion{Impact of the dataset.}
We decided to use the \acrlong{hgd} in our study
even though it was not specifically created to learn embeddings or benchmark embedding approaches. While its high
number of trials per subject (around 1000) allows for the benchmarking of various scenarios, results need to be judged in the light of the mediocre number of 14 subjects in this dataset. Clear advantages of this dataset are that its authors~\cite{schirrmeister2017deep} were successful in training deep models on it, and that it has been used in follow-up publications.

\discussion{Performance of the \acrshort{fbcsp} baseline method.}
We notice that our implementation of \acrshort{fbcsp}
obtains a lower score than what was reported by the authors of \cite{schirrmeister2017deep} on the same dataset.
We qualify this point by reminding that we use  different dataset preprocessings.
We low-pass filter to frequencies up to 40\,Hz before downsampling to 128\,Hz,
while they allowed up to 128\,Hz frequencies before downsampling to 256\,Hz.
In addition, Schirrmeister and colleagues used a larger trial time window (from 0.5\,s to 4\,s relative to trial start, while we have removed the first 0.5\,s of every trial.
In addition, we can not exclude that our implementation of \acrshort{fbcsp} may differ from theirs on some details.

\discussion{Balancing the influence of subject and task classes}
%
The results of the embedding models on the cross-subject training (cf. Result~\ref{result:embed_comp}) revealed a slightly unexpected finding.

Despite the 2D visualizations of the 8-dimensional embedding spaces need to be taken with a grain of salt, we nevertheless were surprised by the 2D visualization obtained for the lexicographic order configuration (c) depicted in Figure~\ref{fig:tsne_all}.
We expected it to be more similar to visualizations
of the weighted lexicographic order (b) than to those of the product order (d), but the opposite was the case. This is a strong indication, that the weighting introduced for (b) has a strong effect.
Specifically, the visualizations of the embedding spaces (c) and (d),
showed a structural hierarchy: while local clusters contained examples from the same class and the same subjects, clusters representing the same subject but different classes were located in their local neighborhood. At further distance, clusters from other subjects were located.
This organization could have been expected if generated by a lexicographic order with the \subjectt being more important than the \targett label but not for configuration (c).


In future work, we will investigate, how the number of task classes compared to the number of subjects influence the resulting embeddings and their preferences to emphasize either similarities between tasks or subjects.

\discussion{Optimization of the weights of the loss components.}
Result~\ref{result:embed_comp} indicates that an embedding function using a loss configuration
that separates both the classes and the subjects may be competitive (despite its average improvement over other configurations was not not statistically significant).
This loss function has weights associated to each of its components which were chosen ad-hoc for our study instead of optimizing them in a hyper-parameters search, which could have improved the performance of the weighted configuration further. For future work, we see a possible benefit for embracing strategies developed in the field of multitask learning in order to improve the configurations weights.

\discussion{Models are related}
Result~\ref{result:confmat} and the confusion matrices describe that the different models mostly committed similar errors. Based on this observation, we think that an ensemble approach, aiming to exploit different strengths of the four embedding configurations to improve the overall classification accuracy would probably not be successful.

\discussion{Application to clinical measurements.}
The \acrlong{hgd} was recorded from healthy subjects only.
As data obtained from patients typically shows more variance both between and within
subjects, it is unclear how well our results may generalize to patient applications. With the weighted lexicographic configuration providing added flexibility by its weighting parameters, this could probably form a good starting point for further research.


\discussion{Feature explicability.}
In our study, we did not try to characterize the intermediate features
extracted by the embedding functions or to investigate their importance by ablation studies or other introspection methods. In future work, it would be interesting to compare what spatial and temporal filters are learned and how the competing approaches differ with respect to these features.


\discussion{Architecture of the embedding function.}
In this study, we chose EEGNet~\cite{lawhern2018eegnet} as an embedding function.
as it currently is one of the most popular and best-performing deep learning architectures for EEG classification.
Lawhern and colleagues showed that it is particularly versatile
with respect to multiple paradigms and obtained similar or better scores than what had been reported for previous deep architectures like ShallowConvNet and DeepConvNet~\cite{schirrmeister2017deep}.
It is also a relatively light architecture because of its restricted number of parameters,
allowing for a relatively fast prototyping in this exploratory study.
To maximize comparability and to reach a fair comparison between the different training procedures, we decided to keep the same architecture (and thus the same expressivity level) for the embedding functions and the deep classification baseline models.
While in this study we did not invest into optimizing the architecture of
the embedding function, it could be interesting to develop an architecture
specific to EEG embeddings in future work.


\discussion{Conclusion}

We have seen, throughout this exploratory study,
that it is possible to learn embeddings for neurophysiological signals
in particular with the novel loss we presented.
We observed that, in the within-subject scenario, using embeddings allowed
us to surpass all the state of the art classification baselines.
We also went beyond ~\cite{alwasiti2020motor} by demonstrating the advantage in using a triplet loss
over a classification loss in the context of a training with few samples.
We presented hybrid classification pipelines,
mixing a deep learning part used to represent the data
and a classical machine learning part to do the classification.
The partial LOSO scenarios truly take advantage of both worlds,
using the generalisation ability of the deep model to generate good cross-subject representations
and the speed of the simple classifiers to have a very fast subject-specific fine-tuning.
The partial LOSO embedding models were able to beat, by a large margin,
the vanilla cross-subject deep classifiers
and reached the within-subject \acrshort{fbcsp} baselines.
In addition to these good scores,
there is still room for improvement of the embedding representation.
Using a dataset thought and recorded specifically to learn embeddings
should lead to a significant improvement.
Also, the loss function we used is not perfect, it can still be optimized or completely remodeled.
Finally, we believe that a true cross-subject and cross-task embedding can be built.
This would lead to a drastic reduction of the amount of data required for new BCI studies.
And it would also allow online applications to have very short to nonexistent calibration times.

\clearpage
\bibliographystyle{unsrtnat}


\clearpage
\appendix
\section{Supplementary materials}
\setcounter{figure}{0}

\subsection{Additional classification results}
We present here some complementary results to those in Section~\ref{sec:results}.

\begin{figure}[h!]
  \centering
  \includegraphics[scale=.5]{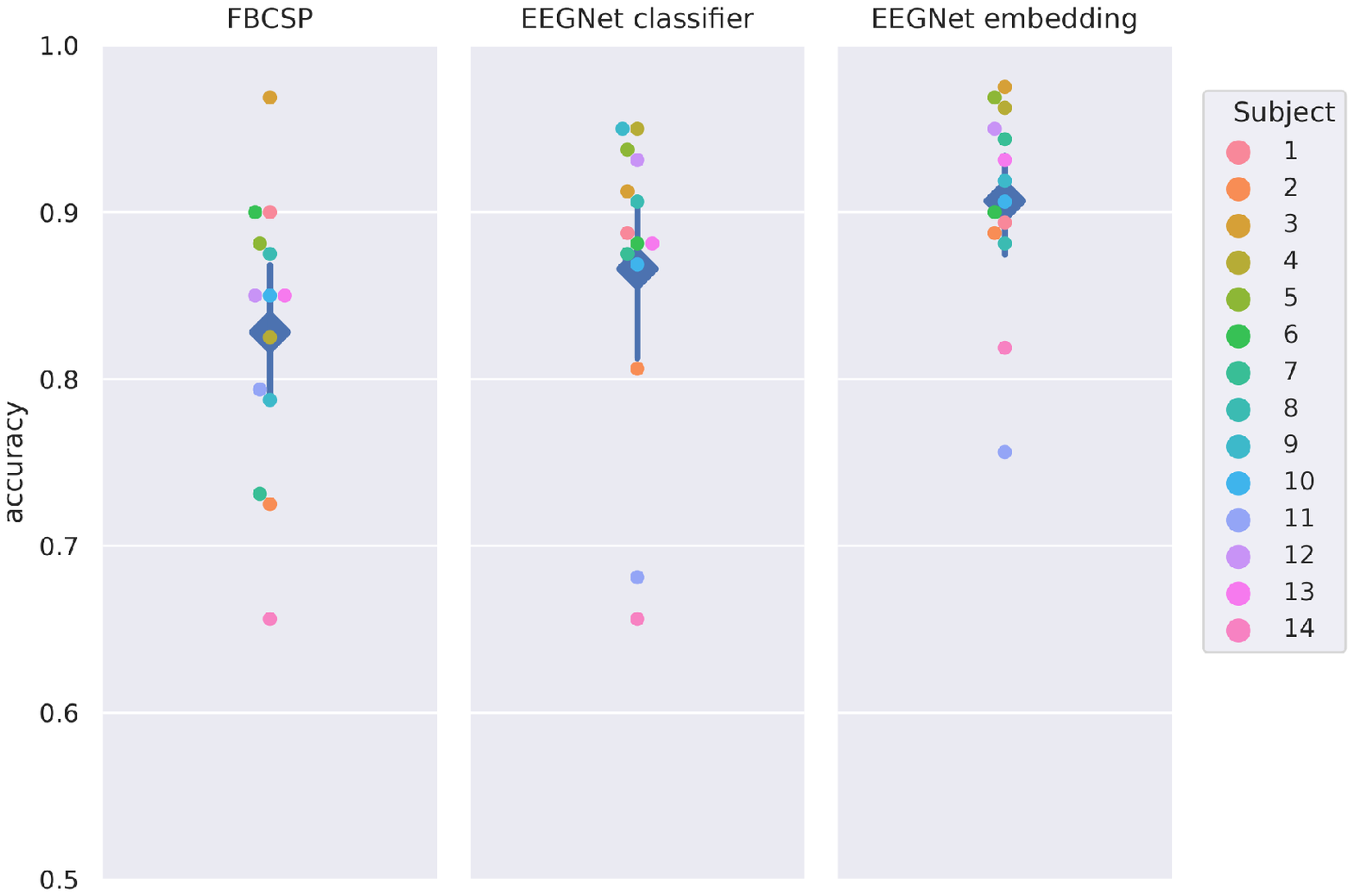}
  \caption[Within-subject classification performance (nearest neighbor classifiers)]
  {Classification performance on \textbf{within-subject} training.
  Each grey box contains the results of a given model.
  The main markers represent the average across subjects and the bars correspond to the standard errors.
  The embedding vectors are classified here by one nearest neighbor classifiers.
  For the results with logistic regression classifiers, see Figure~\ref{fig:acc_singl-subj_logistic}. }
  \label{fig:acc_singl-subj_1nn}
\end{figure}

\begin{figure}[h!]
  \centering
  \includegraphics[scale=.5]{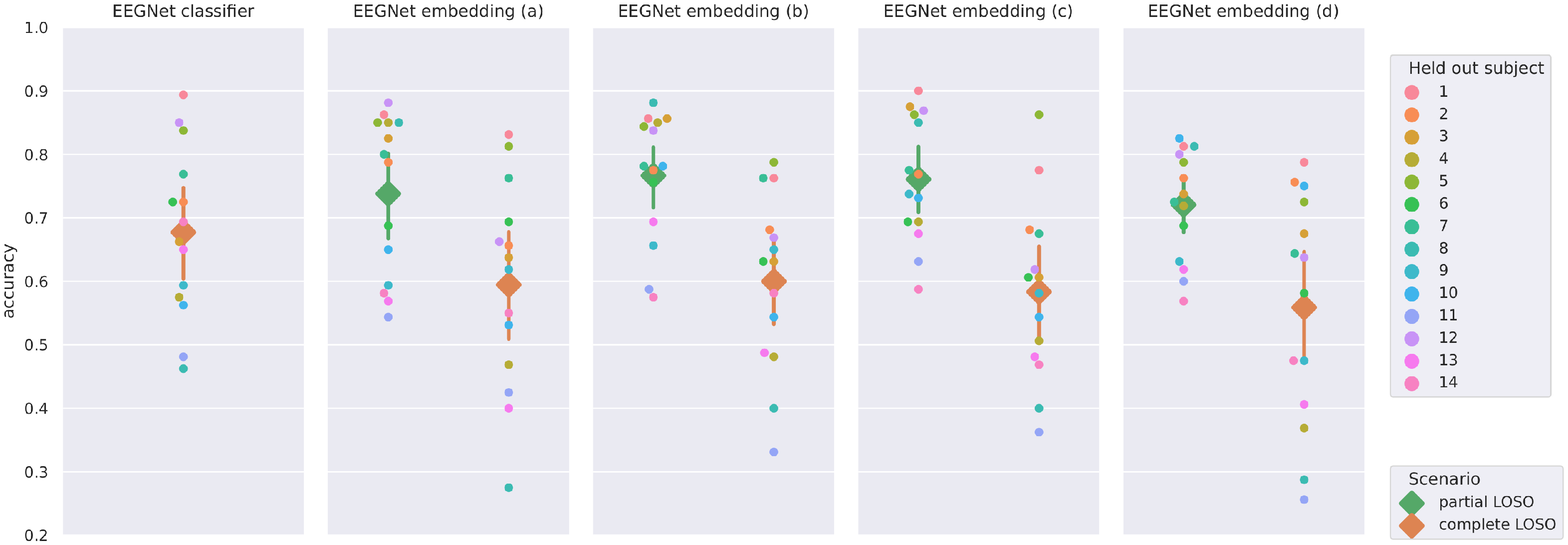}
  \caption[Cross-subject classification performance (nearest neighbor classifiers)]
  {Classification performance on \textbf{cross-subject} training.
  Each grey box contains the results of a given model.
  The boxes of the embedding models each contain two pipelines corresponding to the complete and partial LOSO scenarios.
  The large markers represent the average across subjects and the bars correspond to the standard errors,
  the orange ones are for partial LOSO models and the green ones for complete LOSO.
  The embedding vectors are classified here by one nearest neighbor classifiers.
  For the results with logistic regression classifiers, see Figure~\ref{fig:acc_cross-subj_logistic}. }
  \label{fig:acc_cross-subj_1nn}
\end{figure}
Figures~\ref{fig:acc_singl-subj_1nn} and~\ref{fig:acc_cross-subj_1nn}
present the classification accuracy of the different models.

\begin{figure}[h!]
  \includegraphics[width=\linewidth]{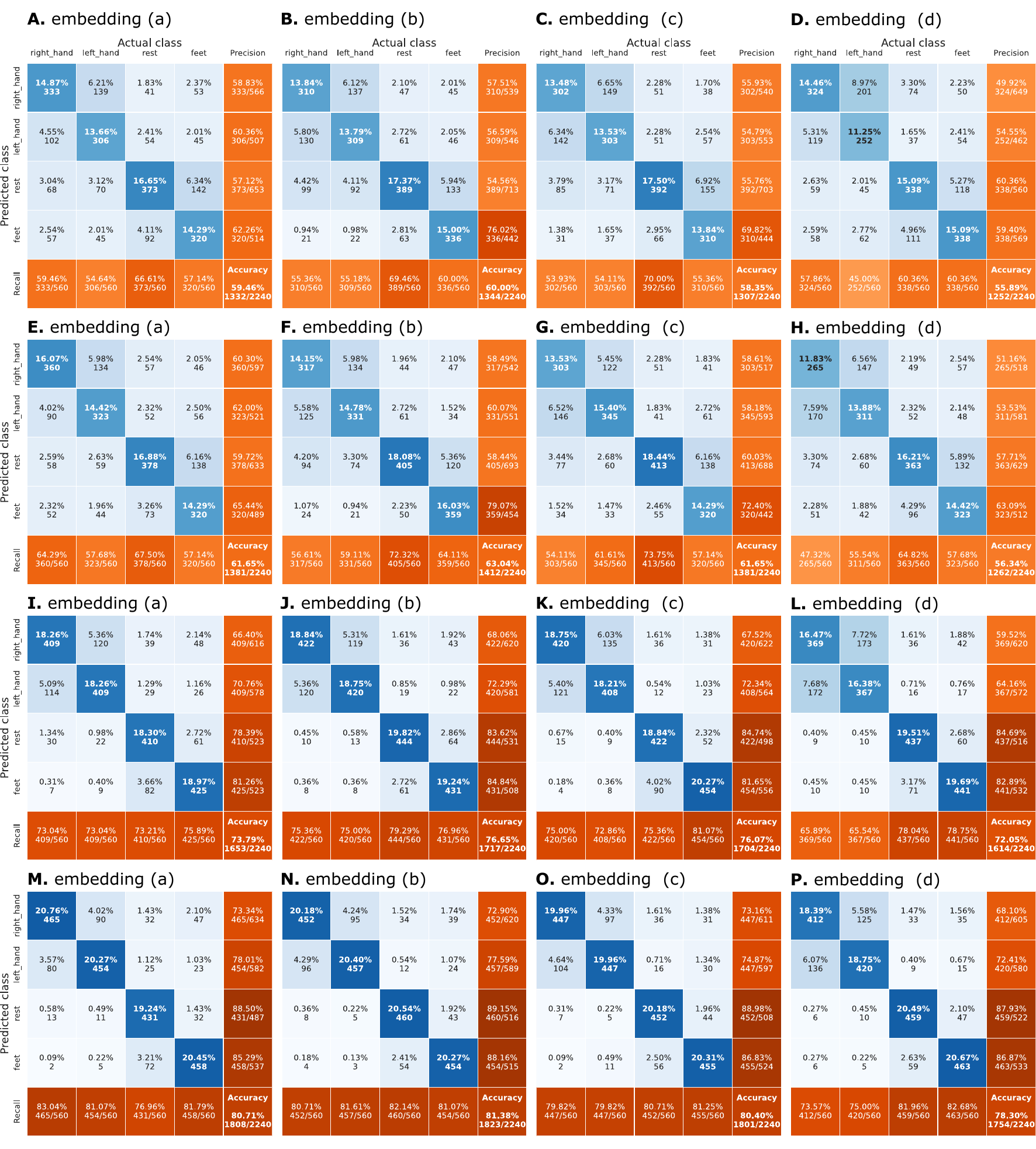}
  \caption[Confusion matrices - embedding models]{Confusion matrices of the embedding models.
    On the first row (subfigures \textbf{A}, \textbf{B}, \textbf{C} and \textbf{D}),
    the embedding vectors were classified by \emph{(one) nearest neighbor} classifiers in \emph{complete LOSO}.
    On the second row (subfigures \textbf{E}, \textbf{F}, \textbf{G} and \textbf{H}),
    the embedding vectors were classified by \emph{logistic regression} in \emph{complete LOSO}.
    On the third row (subfigures \textbf{I}, \textbf{J}, \textbf{K} and \textbf{L}),
    the embedding vectors were classified by \emph{(one) nearest neighbor} classifiers in \emph{partial LOSO}.
    And on the last row (subfigures \textbf{M}, \textbf{N}, \textbf{O} and \textbf{P}),
    the embedding vectors were classified by \emph{logistic regression} in \emph{partial LOSO}.
  }
  \label{fig:confmat_embeddings_appendix}
\end{figure}
Figure~\ref{fig:confmat_embeddings_appendix} presents the confusion matrices of the different classification pipelines
based on embedding functions and trained in cross-subject scenarios (complete and partial LOSO).

\begin{figure}[h!]
  \centering
  \includegraphics[scale=.5]{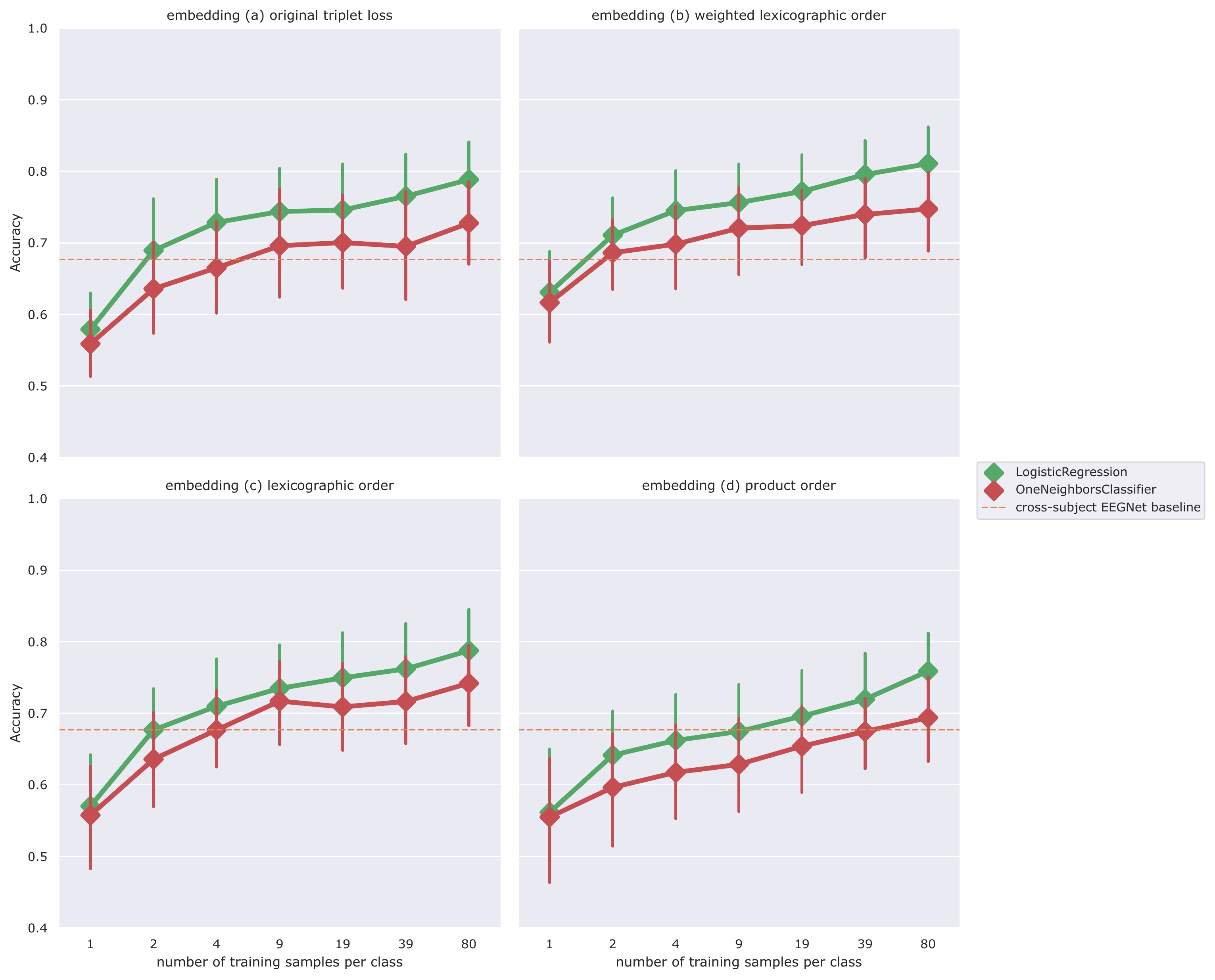}
  \caption[Fine-tuning on few samples - additional results]{
    This Figure contains the results of the cross-subject embedding pipelines trained in partial LOSO
    with different amounts of calibration data (from the test subject).
    Each grey box contains the classification accuracies obtained using one of the loss configuration,
    from (a) on the top left to (d) on the bottom right.
    The scores of the logistic regression classifiers are in green and the (one)-nearest-neighbor classifiers in red.
    Each dot is the average score over the fourteen subjects for a given number of training samples per imagery class and
    the vertical bars represent the standard errors.
    The number of training samples per imagery class can be found on the x-axis,
    going from 1 to 80 samples and evenly distributed on a logarithmic scale.
    When we use $n$ samples for training, these are the $n$ which were chronologically
    recorded the first for that subject.
    We do not go up to 80 training samples because one subject only has that amount
    but most of the other subjects contain 220 training samples per class.
    The performance of the cross-subject EEGNet baseline is
    recalled as the orange doted line
  }
  \label{fig:few_samples_all}
\end{figure}

In Figure~\ref{fig:few_samples_all} are presented complementary materials obtained
from the experiment described in Result~\ref{result:few_samples_cross}.

\subsection{\acrshort{umap} embedding spaces visualisation}
Figures~\ref{fig:umap_all} and~\ref{fig:umap_im-class} show 2D visualizations of the embedding spaces created by \acrshort{umap}.
\embedFigure{umap_all}{2}{\acrshort{umap}}{
  The colors correspond to the subjects
  and the shapes to the imagery classes.\label{fig:umap_all}}
\embedFigure{umap_im-class}{2}{\acrshort{umap}}{
  The colors correspond to the imagery classes.\label{fig:umap_im-class}}

\end{document}